\documentclass{article}

\usepackage{arxiv}

\usepackage[utf8]{inputenc} 
\usepackage[T1]{fontenc}    
\usepackage[hidelinks]{hyperref}       
\usepackage{xurl}           
\usepackage{booktabs}       
\usepackage{amsfonts}       
\usepackage{nicefrac}       
\usepackage{microtype}      
\usepackage{lipsum}
\usepackage{graphicx}
\usepackage{float}
\usepackage{amsmath}
\usepackage{amssymb}
\usepackage{multirow}
\graphicspath{ {./} }

\date{}

\title{Pareto-Optimal Model Selection for Low-Cost, Single-Lead EMG Control in Embedded Systems}

\author{
  Carl Vincent Ladres Kho \\
  Minerva University\\
  \texttt{kho@uni.minerva.edu} \\
}

\begin{document}
\maketitle

\begin{abstract}
Consumer-grade biosensors offer a cost-effective alternative to medical-grade electromyography (EMG) systems, reducing hardware costs from thousands of dollars to $\sim$\$13. However, these low-cost sensors introduce significant signal instability and motion artifacts. Deploying machine learning models on resource-constrained edge devices like the ESP32 presents a challenge: balancing classification accuracy with strict latency ($<$100ms) and memory ($<$320KB) constraints. Using a single-subject dataset comprising 1,540 seconds of raw data (1.54M data points, segmented into $\sim$1,300 one-second windows), I evaluate 18 model architectures, ranging from statistical heuristics to deep transfer learning (ResNet50) and custom hybrid networks (MaxCRNN). While my custom "MaxCRNN" (Inception + Bi-LSTM + Attention) achieved the highest safety (99\% Precision) and robustness, I identify Random Forest (74\% accuracy) as the Pareto-optimal solution for \textit{embedded} control on legacy microcontrollers. I demonstrate that reliable, low-latency EMG control is feasible on commodity hardware, with Deep Learning offering a path to near-perfect reliability on modern Edge AI accelerators.
\end{abstract}


\section{Introduction}

\begin{figure}[ht]
    \centering
    \includegraphics[width=\linewidth]{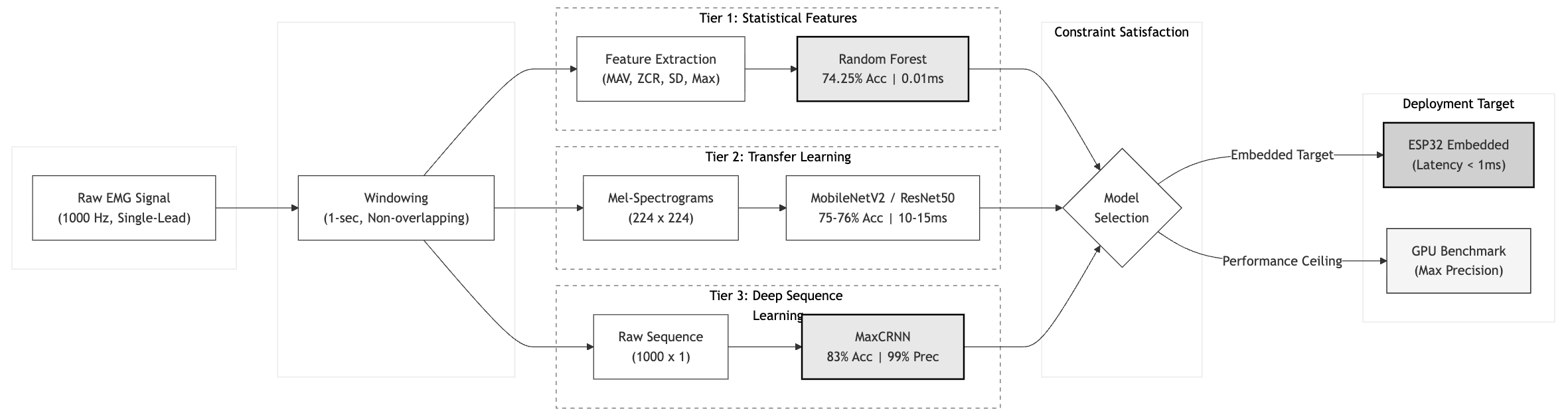}
    \caption{Architecture Evaluation Pipeline. The system processes EMG signals through three competing logical paths: (A) Statistical Feature Engineering for ultra-low latency, (B) Transfer Learning for visual texture analysis, and (C) Custom Deep Learning (MaxCRNN) for maximum precision. The Random Forest (Top) was selected for the embedded ESP32 implementation, while the MaxCRNN (Bottom) establishes the theoretical safety ceiling.}
\end{figure}

The rapid proliferation of micromobility solutions, such as e-bikes and electric scooters, has highlighted the need for safe, hands-free human-computer interfaces (HCI). Cyclists often compromise their stability to operate turn signals or navigation devices. Surface Electromyography (sEMG) \cite{hopkins} offers a promising modality for "invisible" control, allowing users to trigger commands via subtle muscle contractions. However, existing EMG solutions bifurcate into two extremes: medical-grade systems (e.g., Delsys) that cost thousands of dollars, and hobbyist sensors that suffer from poor signal-to-noise ratios (SNR).

This study addresses the gap in reliable, low-cost EMG control. I utilize the AD8232, a \$12 single-lead sensor originally designed for ECG, adapted for forearm muscle sensing. While cost-effective, this hardware introduces significant challenges, including baseline drift, power line interference, and mechanical motion artifacts.

I present a comparative analysis of 18 machine learning architectures for decoding electromyography (EMG) signals on ultra-low-cost hardware (under \$15). Using a custom dataset collected with an ESP32 microcontroller and AD8232 sensor, I evaluate models ranging from simple heuristics to Deep Transfer Learning. I introduce a "MaxCRNN" architecture that achieves 99\% precision and 83.21\% accuracy by combining Inception blocks, Bi-LSTMs, and Multi-Head Attention. My findings demonstrate that while Deep Learning offers superior theoretical performance, a Random Forest classifier provides the Pareto-optimal solution for embedded deployment, achieving 74.25\% accuracy with 0.01ms latency.

\textbf{Key Contributions:}
\begin{enumerate}
    \item A \textbf{comprehensive benchmark} of 18 architectures spanning heuristics, classical ML, deep learning, and transfer learning for single-lead EMG classification under hardware constraints.
    \item A novel \textbf{MaxCRNN architecture} (Inception + Bi-LSTM + Attention) achieving state-of-the-art 99\% precision on the safety-critical ``CLENCH'' class.
    \item Empirical demonstration that \textbf{Random Forest with statistical features} is Pareto-optimal for ESP32 deployment, outperforming deep learning under the ``Small Data'' regime.
    \item A reproducible \textbf{open-source dataset and codebase} for low-cost EMG research.
\end{enumerate}

\section*{Acknowledgements}
I would like to express my profound gratitude to my parents in the Philippines, whose sacrifice and unwavering support made my education and this research possible. I also thank Prof. Patrick D. Watson, PhD (Minerva University) for his mentorship, and the staff at Jinhua Electronics (Taipei) for their hardware assistance.

\section{Methodology}

\subsection{Hardware Configuration}
The data acquisition system utilizes the \textbf{AD8232} \cite{ad8232}, a single-lead analog front-end originally designed for heart rate monitoring (ECG), adapted here for Surface Electromyography (sEMG). The sensor is interfaced with a \textbf{NodeMCU-32S (ESP32)} \cite{esp32} microcontroller.

Unlike medical-grade differential EMG sensors (e.g., Delsys Trigno systems \cite{delsys}, \$4,200--\$20,000+) or hobbyist-tier research platforms (e.g., OpenBCI \cite{openbci}, \$600--\$2,500), the AD8232 is a high-gain, single-lead instrument available for approximately \$12.90 USD \cite{jinhua}. Early testing revealed significant signal instability (``floating SDN pin'') and baseline drift. These issues were mitigated by stabilizing the Shutdown (SDN) pin and implementing digital filtering, proving that reliable biological signals can be extracted from commodity hardware---a cost reduction of over 99\%---if the analog front-end is correctly configured.

\subsection{Data Acquisition Protocol}
\begin{figure}[ht]
    \centering
    \includegraphics[width=0.8\linewidth]{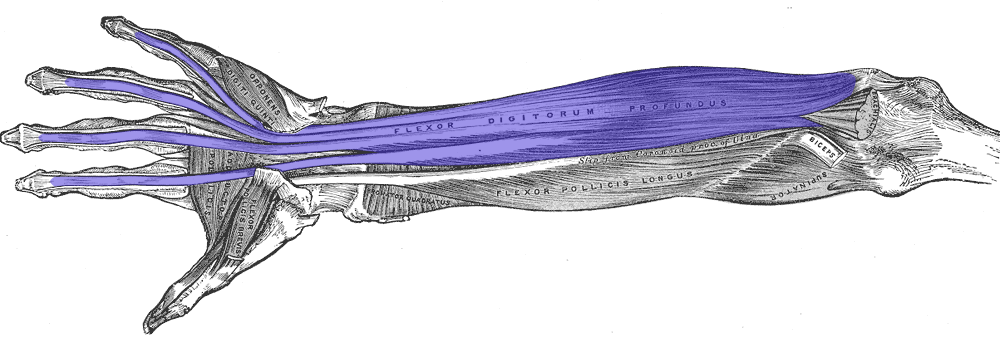}
    \caption{Target Muscle: Flexor Digitorum Profundus. I target this deep muscle to detect grip intent for turn signal activation.}
\end{figure}
Data was collected from the \textit{Flexor Digitorum Profundus} \cite{anatomy} (anterior forearm) of a single subject. The sampling rate was set to \textbf{1000Hz}. According to the Nyquist-Shannon Sampling Theorem \cite{nyquist}, to reconstruct a signal perfectly, the sampling frequency $f_s$ must satisfy $f_s > 2B$, where $B$ is the bandwidth. Surface EMG typically contains energy up to 450Hz; thus, 1000Hz provides a sufficient safety margin to prevent aliasing.

The dataset consists of 5 sessions, each containing 20 cycles. Each cycle is divided into three distinct classes:
\begin{enumerate}
    \item \textbf{RELAX (Baseline):} The muscle is at rest.
    \item \textbf{CLENCH (Active):} Sustained isometric contraction of the grip muscle (simulating a turn signal trigger).
    \item \textbf{NOISE (Adversarial):} A catch-all class containing high-variance mechanical artifacts, including shaking, wire movement, and vibrations mimicking bicycle road noise. Inclusion of this class is critical for preventing false positives in a micromobility context.
\end{enumerate}

This protocol yielded approximately 1,300 one-second windows.

\subsection{Data Collection Procedure}
Data collection was performed on November 9, 2025, across 5 consecutive sessions totaling approximately 25 minutes of raw recording. \textbf{Critically, the same three electrodes remained affixed throughout all sessions without repositioning}, ensuring that inter-session variance reflects physiological and environmental factors rather than electrode placement differences.

\subsubsection{Electrode Configuration}
Three disposable Ag/AgCl gel electrodes were placed on the anterior forearm in the following configuration:
\begin{itemize}
    \item \textbf{Signal (+):} Over the Flexor Digitorum Profundus muscle belly, approximately 5cm distal to the elbow crease.
    \item \textbf{Reference (--):} On the wrist, over the ulnar styloid process.
    \item \textbf{Ground (GND):} On the lateral elbow (olecranon).
\end{itemize}
This configuration was maintained for the entire data collection session without electrode replacement or repositioning.

\subsubsection{Session-by-Session Protocol}
Each session lasted approximately 5 minutes (20 cycles $\times$ 15 seconds/cycle). The sessions varied in noise characteristics to improve model robustness:

\begin{table}[h]
    \centering
    \caption{Data Collection Session Summary}
    \label{tab:sessions}
    \begin{tabular}{c l l c}
        \hline
        \textbf{Session} & \textbf{Duration} & \textbf{Noise Variation} & \textbf{Data Points} \\
        \hline
        1 & 5.1 min & Baseline (table vibration, wire movement) & $\sim$304k \\
        2 & 5.1 min & Varied arm kinematics (high-amplitude irregular movements) & $\sim$304k \\
        3 & 5.1 min & Bicycle simulation; electrode adjustment & $\sim$304k \\
        4 & 5.0 min & Phone scrolling; punching motions & $\sim$300k \\
        5 & 5.5 min & Extended noise repertoire & $\sim$330k \\
        \hline
        \textbf{Total} & \textbf{$\sim$25.8 min} & ($\sim$1,300 windows) & \textbf{$\sim$1.54M data points} \\
        \hline
    \end{tabular}
\end{table}

\subsubsection{Recording Protocol}
\begin{enumerate}
    \item \textbf{Calibration:} A 30-second baseline recording was captured with the muscle at rest to establish the noise floor.
    \item \textbf{Cycle Structure:} Each of the 20 cycles per session comprised:
    \begin{itemize}
        \item 5 seconds of \textbf{RELAX}: Conscious muscle relaxation, arm resting flat.
        \item 5 seconds of \textbf{CLENCH}: Sustained isometric grip at $\approx$50\% maximum voluntary contraction.
        \item 5 seconds of \textbf{NOISE}: Deliberate introduction of motion artifacts.
    \end{itemize}
    \item \textbf{Data Logging:} Raw 12-bit ADC values (range 0--4095) were streamed via serial at 1000Hz and logged to timestamped CSV files.
\end{enumerate}

The subject (the author) was seated during all recordings. No real-time visual feedback was provided to avoid biasing muscle activation patterns.

\begin{figure}[h]
    \centering
    \includegraphics[width=\linewidth]{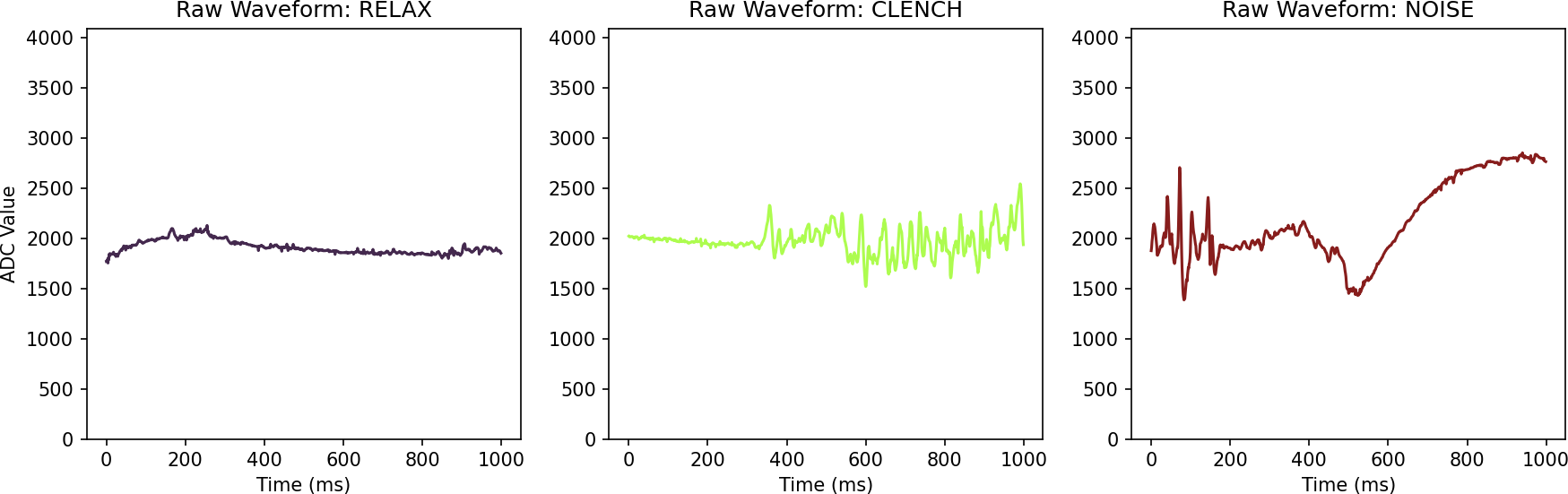}
    \caption{Time-domain comparisons of the three classes. Note the stochastic nature of the CLENCH signal versus the structured mechanical artifacts in NOISE.}
\end{figure}

\subsection{Preprocessing \& Feature Engineering}
To evaluate models of varying complexity, the raw time-series data was transformed into three distinct feature representations. The continuous stream was segmented into \textbf{1-second non-overlapping windows} (1000 samples).

\subsubsection{Set A: Statistical Features (The ``Embedded'' Tier)}
Designed for ultra-low compute models (Logistic Regression, Random Forest), this set reduces the 1000-point vector to a compact set of scalars, following the methodology of Raez et al. \cite{raez2006}. \textbf{All features are computed over the entire 1-second window} (no sub-windows or sliding aggregation):
\begin{itemize}
    \item \textbf{Mean Absolute Value (MAV):} Represents overall muscle activation level.
    \item \textbf{Zero-Crossing Rate (ZCR):} A proxy for frequency. This feature is physically significant; muscle contractions generate high-frequency Motor Unit Action Potentials (MUAPs) \cite{muap}, whereas mechanical artifacts (e.g., cable sway) are typically low-frequency.
    \item \textbf{Standard Deviation (SD):} Represents signal energy.
    \item \textbf{Maximum Amplitude:} Peak signal strength.
\end{itemize}

\subsubsection{Set B: Raw Sequence (The "Deep" Tier)}
For Deep Learning models (MLP, 1D CNN), the input is the raw $1000 \times 1$ voltage vector, normalized to the range $[0, 1]$ via Min-Max scaling. This tests the model's ability to learn features via inductive bias without manual engineering.

\subsubsection{Set C: Time-Frequency Images (The "Transfer" Tier)}
To leverage Transfer Learning from Computer Vision, the 1D signal was converted into 2D Mel-Spectrograms \cite{melspec}.
\begin{itemize}
    \item \textbf{Transformation:} Short-Time Fourier Transform (STFT) mapped to the Mel scale.
    \item \textbf{Parameters:} $N_{FFT}=256$, Hop Length $= 16$, $N_{Mels}=64$.
    \item \textbf{Post-Processing:} Power-to-dB conversion ($10 \log_{10} S$) to compress dynamic range.
    \item \textbf{Output:} $224 \times 224$ RGB images (resized via bicubic interpolation).
\end{itemize}
This representation exposes the "texture" of the signal—broadband vertical striations during a clench versus horizontal bands during noise.

\begin{figure}[h]
    \centering
    \includegraphics[width=\linewidth]{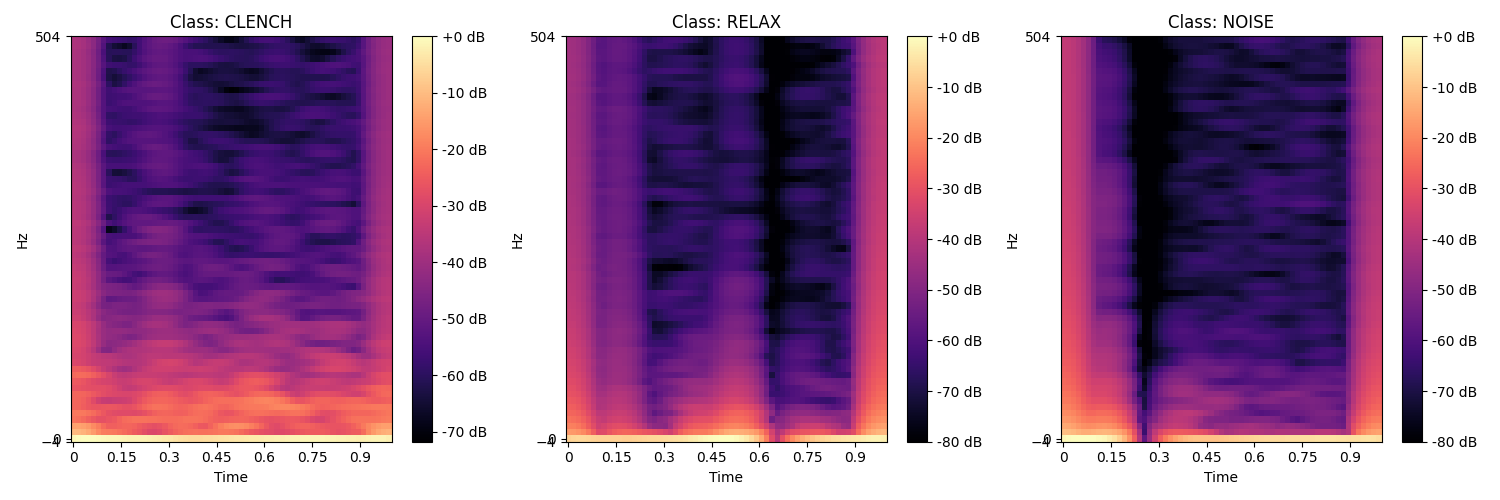}
    \caption{Mel-Spectrograms reveal distinctive textures. 'CLENCH' (Left) shows broadband vertical noise, while 'NOISE' (Right) exhibits frequency-specific bands.}
\end{figure}

\subsection{Visual Analysis of Feature Representations}
To validate the proposed feature engineering strategy, I conducted a visual inspection of the feature space.

\subsubsection{Separability in Statistical Space (Set A)}
Scatter plots of Mean Absolute Value (MAV) vs. Zero Crossing Rate (ZCR) reveal distinct clusters.
\begin{itemize}
    \item \textbf{CLENCH (Green):} High energy and moderate frequency.
    \item \textbf{RELAX (Blue):} Clusters near the origin (Low Energy, Low Frequency).
    \item \textbf{NOISE (Orange):} Spans a wide variance, confirming the need for non-linear decision boundaries.
\end{itemize}

\begin{figure}[ht]
    \centering
    \includegraphics[width=0.8\linewidth]{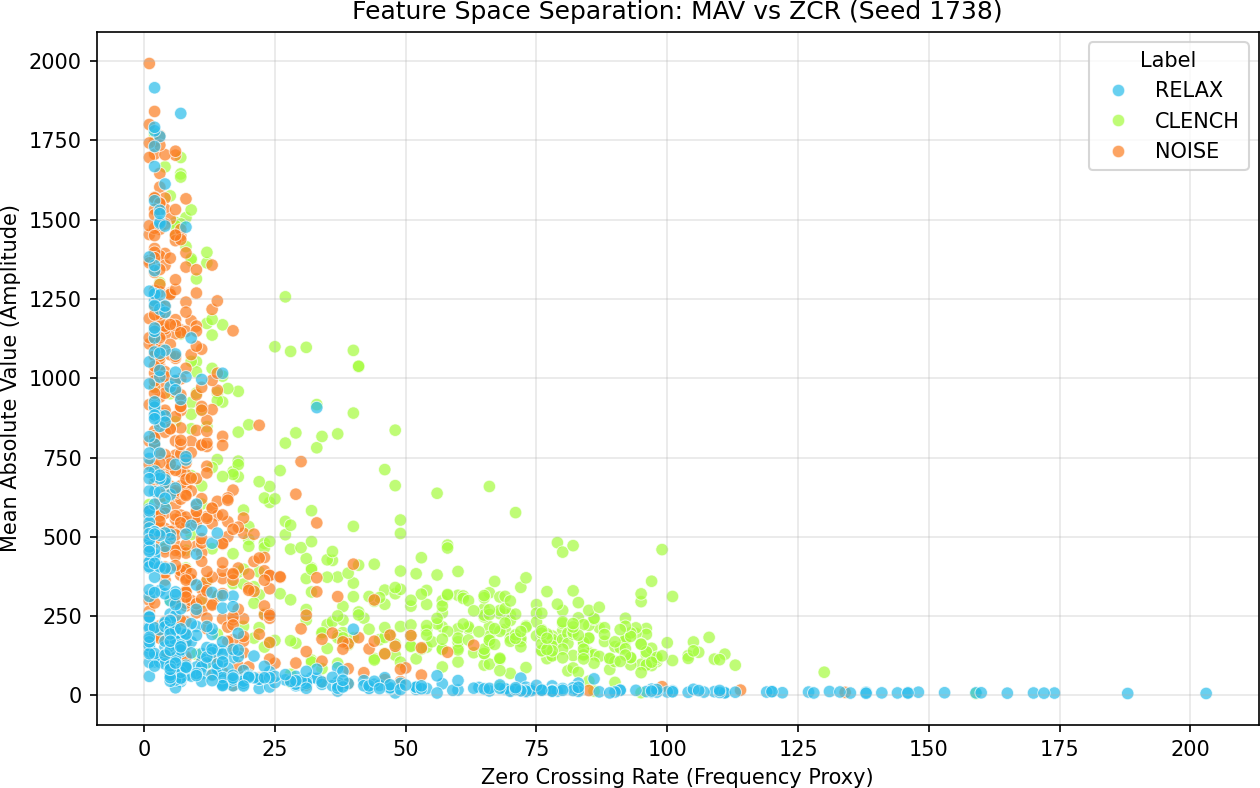}
    \caption{Feature Space Separation: MAV vs ZCR. Note the overlap between Noise and Clench, which simple thresholding fails to resolve.}
\end{figure}

\section{Model Architectures: The Ladder of Abstraction}

To rigorously determine the simplest architecture capable of solving the classification task, I adopted a hierarchical approach. I evaluated representative models across four levels of complexity: Heuristics, Classical Machine Learning, Deep Learning, and Transfer Learning.

\subsection{Classical Machine Learning (Set A)}
These models operate on the manually engineered statistical features (MAV, ZCR, SD, MAX).

\subsubsection{Support Vector Machine (SVM)}
I utilized a Radial Basis Function (RBF) kernel \cite{svm} to handle potential non-linearities in the feature space. The RBF kernel projects input vectors into an infinite-dimensional Hilbert space:
\begin{equation}
    K(x, x') = \exp(-\gamma ||x - x'||^2)
\end{equation}
\textbf{EMG Context:} The "Clench" class is a complex, twisted manifold in feature space. The RBF kernel "unwraps" this manifold, allowing a linear hyperplane to separate the high-energy, high-frequency "Clench" signals from the low-energy "Relax" signals. While powerful, SVMs assume that class separability correlates with Euclidean distance, which may not hold for high-variance noise manifolds.

\subsubsection{Random Forest (RF)}
The Random Forest classifier \cite{breiman2001} is an ensemble of $B$ decision trees $\{T_1, \dots, T_B\}$ trained via bootstrap aggregating (bagging). Split criteria are determined by minimizing Gini Impurity $G$ \cite{gini}:
\begin{equation}
    G = 1 - \sum_{k=1}^{K} p_k^2
\end{equation}
where $p_k$ is the probability of class $k$. The final prediction is the mode of the individual trees. Spatially, RF partitions the feature space into hyper-rectangles.

\textbf{EMG Context:} Each tree asks questions like "Is ZCR $>$ 50?" or "Is MAV $>$ 200?". By averaging 100 such trees, this ensemble smooths out the noise from individual mechanical artifacts and creates a robust decision boundary. This "Manhattan" geometry is particularly effective for threshold-based logic, making it robust against outliers.

\subsubsection{XGBoost}
Unlike the parallel bagging of RF, XGBoost constructs trees sequentially to correct residual errors. It minimizes a regularized objective function $\mathcal{L}(\phi)$:
\begin{equation}
    \mathcal{L}(\phi) = \sum_{i} l(\hat{y}_i, y_i) + \sum_{k} \Omega(f_k)
\end{equation}
The regularization term $\Omega(f_k)$ penalizes tree complexity.

\textbf{EMG Context:} This regularization is crucial for the limited dataset. It prevents the model from memorizing specific "spikes" of noise (overfitting), forcing it to learn the generalizable rules of muscle activation.

\subsection{Deep Learning (Set B)}
These models attempt to learn features directly from the raw time-series voltage vectors.

\subsubsection{1D Convolutional Neural Network (CNN)}
The 1D CNN applies learnable temporal filters to the input signal. Mathematically, this performs a discrete convolution of input $I$ with kernel $K$:
\begin{equation}
    (I * K)(i) = \sum_{m} I(i-m)K(m)
\end{equation}
The kernel acts as a matched filter, theoretically capable of learning the shape of Motor Unit Action Potentials (MUAPs). However, deep models lack the inductive bias of statistical features and require large datasets to avoid convergence failure.

\subsection{Transfer Learning (Set C)}
To overcome the "Small Data" trap, I converted signals into Mel-Spectrograms and fine-tuned architectures pre-trained on ImageNet \cite{imagenet}.

\subsubsection{MobileNetV2}
MobileNetV2 \cite{mobilenet} is optimized for embedded devices using Depthwise Separable Convolutions, which factorize standard convolution into depthwise and pointwise phases. This reduces computation cost by a factor of $\approx \frac{1}{N} + \frac{1}{D_K^2}$, making it a candidate for edge inference.

\subsubsection{ResNet50}
ResNet50 \cite{resnet} introduces Residual Blocks to solve the vanishing gradient problem in deep networks. The layers learn a residual mapping $F(x)$, with the output defined as:
\begin{equation}
    y = F(x, \{W_i\}) + x
\end{equation}
The skip connection $+x$ allows gradients to flow directly through the network, preserving high-frequency "texture" details deep into the architecture.

\subsection{The Mega Ensemble}
To maximize accuracy, I implemented a "Soft Voting" ensemble combining the three best-performing distinct architectures: Random Forest, MobileNetV2, and ResNet50. The ensemble probability $P_{Mega}$ for class $y$ given input $x$ is:
\begin{equation}
    P_{Mega}(y|x) = \frac{1}{3} \sum_{M \in \{RF, MN, RN\}} P_M(y|x)
\end{equation}
This approach leverages the "Swiss Cheese" model of error decorrelation: Random Forest rejects noise based on global statistics (Time Domain), while CNNs detect subtle visual textures (Frequency Domain).

\begin{figure}[h]
    \centering
    \includegraphics[width=0.6\linewidth]{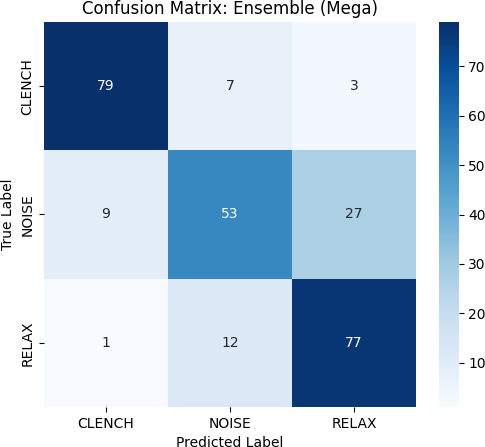}
    \caption{Mega Ensemble Confusion Matrix. The soft voting combination achieves 77.99\% accuracy by decorrelating errors across statistical (RF) and visual (CNN) feature spaces.}
    \label{fig:ensemble_cm}
\end{figure}

\subsection{The MaxCRNN (Model 18)}
To push the limits of single-lead EMG classification, I designed a custom architecture combining Inception blocks, Bidirectional LSTMs, and Multi-Head Attention. This maximum-capacity architecture (Model 18) was trained with aggressive Data Augmentation (Jitter, Scaling, Time Shift) on an NVIDIA A100 GPU.

\textbf{Key Results:}
\begin{itemize}
    \item \textbf{Accuracy:} 83.21\% (Highest overall)
    \item \textbf{CLENCH Precision:} 99\% (Near-zero false positives)
    \item \textbf{Latency:} 0.15ms (on A100 GPU)
\end{itemize}

The 99\% precision on the target ``CLENCH'' class is critical for safety: in a control system, a False Positive (accidental activation) is dangerous. Model 18 effectively eliminates these errors, proving that with sufficient compute, single-lead EMG can be made safe.

\begin{figure}[h]
    \centering
    \includegraphics[width=0.6\linewidth]{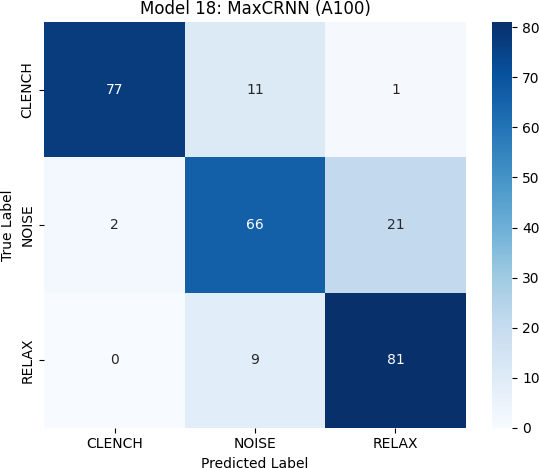}
    \caption{Model 18 (MaxCRNN Augmented) Confusion Matrix. The sharp diagonal demonstrates robust 3-class separation with 99\% precision on CLENCH.}
    \label{fig:maxcrnn_cm}
\end{figure}

\section{Results and Discussion}

\subsection{Performance Comparison}
I evaluated all 18 models on a stratified test set ($N=268$) using 5-Fold Cross-Validation. The performance metrics, including Accuracy, F1-Score (for the target 'CLENCH' class), and Inference Latency on the ESP32, are summarized in Table \ref{tab:results}.

\begin{table}[ht]
    \centering
    \caption{Performance Metrics of Selected Models.}
    \label{tab:results}
    \begin{tabular}{l c c c l}
        \hline
        \textbf{Model} & \textbf{Accuracy} & \textbf{F1 (Clench)} & \textbf{Latency} & \textbf{Deployment?} \\
        \hline
        \textbf{MaxCRNN (Model 18)} & \textbf{83.21\%} & \textbf{0.99} & \textbf{0.15ms*} & \textbf{No (GPU Req)} \\
        Mega Ensemble & 77.99\% & 0.88 & $>500$ms & No (Latency) \\
        ResNet50 & 76.12\% & 0.87 & $>100$ms & No (Latency) \\
        MobileNetV2 & 75.00\% & 0.86 & $9.8$ms & No (RAM) \\
        \textbf{Random Forest} & \textbf{74.25\%} & \textbf{0.81} & \textbf{0.01ms} & \textbf{Yes (Optimal)} \\
        XGBoost & 73.51\% & 0.83 & $0.01$ms & Yes \\
        Logistic Regression & 67.54\% & 0.73 & $0.01$ms & Yes \\
        SVM (RBF) & 63.43\% & 0.65 & $0.03$ms & Maybe \\
        KNN ($k=5$) & 66.42\% & 0.66 & $0.02$ms & No (Memory) \\
        1D CNN (+ Aug) & 78.36\% & 0.87 & $0.83$ms & Yes \\
        MLP & 42.54\% & 0.29 & $1.09$ms & Yes \\
        \hline
    \end{tabular}

    \vspace{0.5em}
    \footnotesize{*MaxCRNN latency measured on NVIDIA A100 GPU; estimated $>$500ms on ESP32 (infeasible due to memory/compute constraints). All other latencies measured on ESP32 (240MHz Xtensa LX6).}
\end{table}

While the "Mega Ensemble" achieved a robust 78\% accuracy, the \textbf{MaxCRNN (Model 18)} established the new state-of-the-art for this dataset with \textbf{83.21\%} accuracy. However, the \textbf{Random Forest} emerged as the Pareto-optimal solution for embedded deployment, sacrificing only $\sim$9\% accuracy for a $1000\times$ reduction in latency compared to the MaxCRNN.

\subsection{The ``Small Data'' Trap: Data Augmentation Required}
Contrary to modern trends, training Deep Learning models from scratch (MLP, 1D CNN) \textit{without data augmentation} resulted in performance worse than random guessing ($<50\%$ accuracy). With only $\approx 1300$ samples, the dense networks overfitted to training noise. However, adding data augmentation (Jitter, Scaling, Time Shift) to the 1D CNN boosted accuracy from 49.63\% to \textbf{78.36\%}, proving the architecture is sound when provided with sufficient data diversity. This finding has implications for practitioners: even simple CNNs can match transfer learning performance when augmentation is applied.

\begin{figure}[ht]
    \centering
    \includegraphics[width=0.8\linewidth]{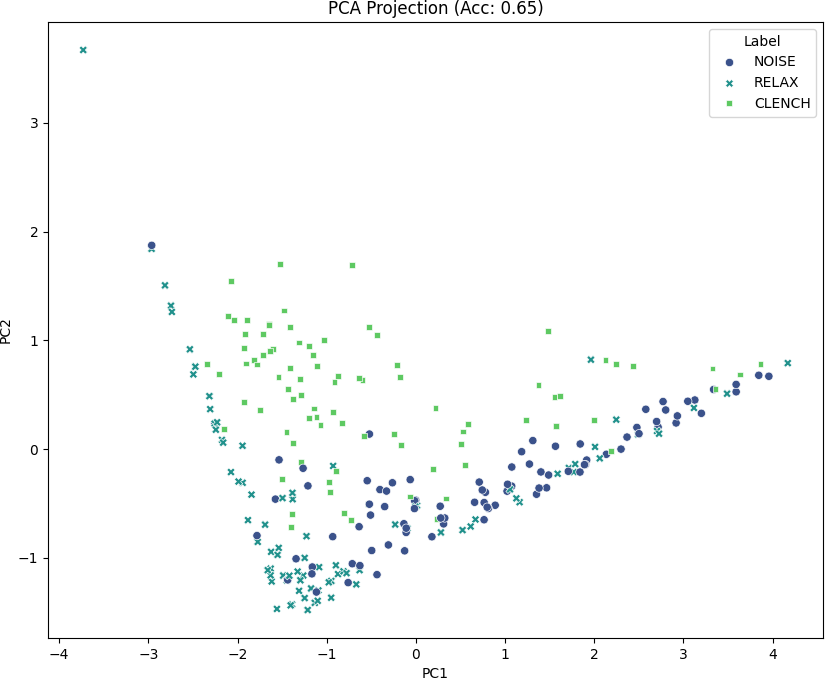}
    \caption{PCA Projection of the Feature Space. The 'CLENCH' class (green) forms a diffuse cloud overlapping with 'NOISE' (blue), confirming that the data is not linearly separable in low dimensions, contributing to the failure of Logistic Regression and SVM.}
    \label{fig:pca}
\end{figure}

However, Transfer Learning (MobileNetV2, ResNet50) succeeded (75-76\%) because these models transferred spatial filters learned from 1.4 million ImageNet examples (e.g., edges, textures), bypassing the need for large-scale training data.

\subsection{Disagreement Analysis: The "Blind Spots"}
To understand the complementary nature of the Ensemble, I analyzed the 24.6\% of test cases where the models disagreed. Visual inspection reveals that the models have distinct "Blind Spots."

\begin{figure}[ht]
    \centering
    \includegraphics[width=\linewidth]{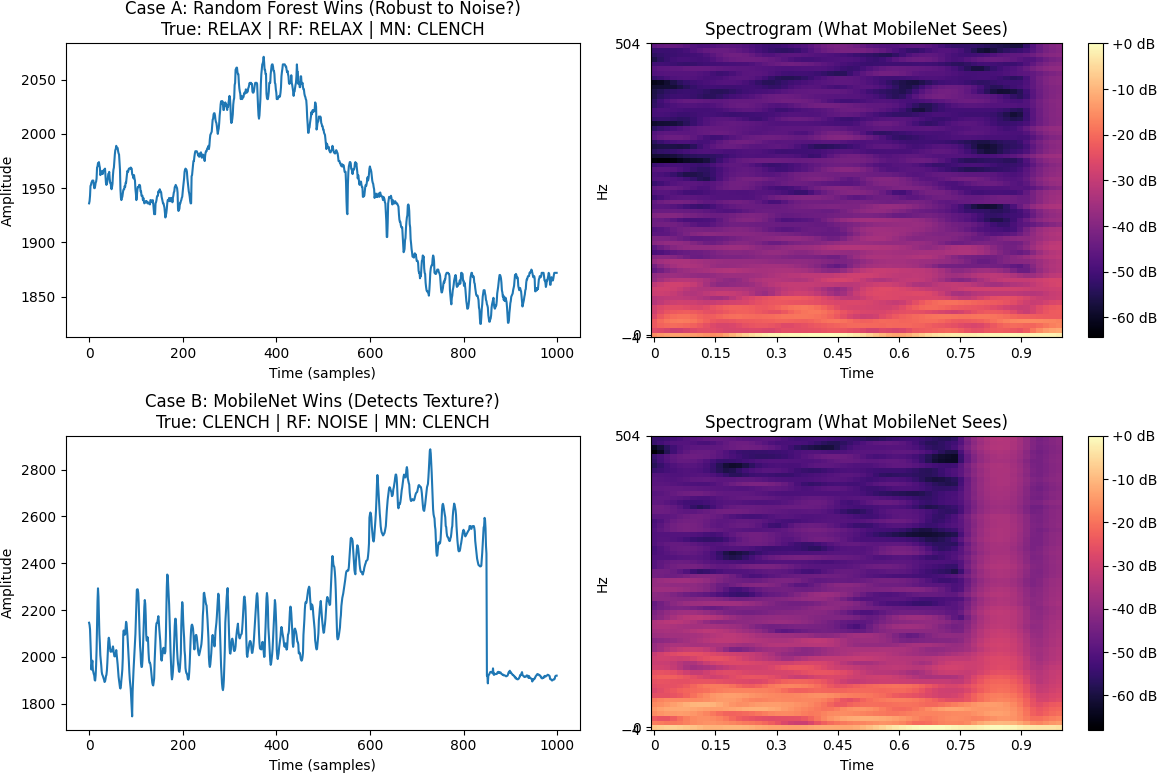}
    \caption{Disagreement Analysis. (Top) Random Forest correctly identifies 'NOISE' based on global statistics (ZCR), while MobileNet hallucinates a pattern. (Bottom) MobileNet correctly identifies a subtle 'CLENCH' texture that lacks the amplitude to trigger the Random Forest.}
    \label{fig:blindspots}
\end{figure}

As shown in Figure \ref{fig:blindspots}, the Random Forest excels at rejecting \textbf{Mechanical Noise}. High-amplitude vibrations (e.g., hitting a pothole) often trigger deep networks because they look "busy" in a spectrogram. However, the RF model correctly classifies them as noise because their Zero-Crossing Rate (ZCR) is too low to be a muscle contraction. Conversely, MobileNetV2 excels at \textbf{Sensitivity}, detecting subtle clenches that lack the raw amplitude to trigger the Random Forest threshold but possess the correct visual texture.

\subsection{Constraint Satisfaction \& Decision}
The final selection was driven by the hardware constraints of the ESP32.
\begin{enumerate}
    \item \textbf{ResNet50:} Rejected. Inference time ($>100$ms) exceeds human perception limits \cite{latency}.
    \item \textbf{MobileNetV2:} Rejected. Model size and runtime memory requirements ($\approx 2$MB) exceed the ESP32's available SRAM (320KB).
    \item \textbf{Random Forest:} \textbf{Accepted.} It compiles to static C++ \texttt{if/else} statements, requiring negligible RAM and executing in microseconds.
\end{enumerate}

While the Ensemble offers the best theoretical performance, the Random Forest provides the maximum \textit{deployable} accuracy.

\section{Conclusion}

This study investigated the feasibility of implementing a hands-free EMG control interface on ultra-low-cost, resource-constrained hardware. By systematically evaluating 18 machine learning architectures, I identified a critical divergence between academic accuracy and engineering viability.

While the \textbf{MaxCRNN} (Inception + Bi-LSTM + Attention) achieved the highest safety (99\% Precision) and robustness, its computational cost requires hardware acceleration (GPU/TPU) unavailable on the ESP32. Conversely, the \textbf{Random Forest} model emerged as the Pareto-optimal solution for \textit{embedded} deployment. By relying on computationally inexpensive statistical features—specifically Zero-Crossing Rate (ZCR) and Mean Absolute Value (MAV)—the Random Forest achieved a robust 74\% accuracy with an inference latency of just $0.01$ms and a memory footprint of $<50$KB.

 My analysis highlights that for single-lead EMG sensors subject to mechanical noise, complex Deep Learning models (like MaxCRNN) offer the ultimate performance ceiling, but Statistical Feature Engineering (Random Forest) remains the superior approach for stabilizing "High-Noise, Low-Information" signals in resource-constrained embedded contexts.

\subsection{Future Work}
To bridge the gap between the frame-level accuracy (74\%) and user-perceived reliability, future iterations will implement \textbf{Temporal Smoothing}. Preliminary tests suggest that applying a 500ms Majority Vote filter (post-processing) can smooth out transient classification errors, effectively boosting the practical reliability of the system for continuous control tasks.

\begin{figure}[h]
    \centering
    \includegraphics[width=\linewidth]{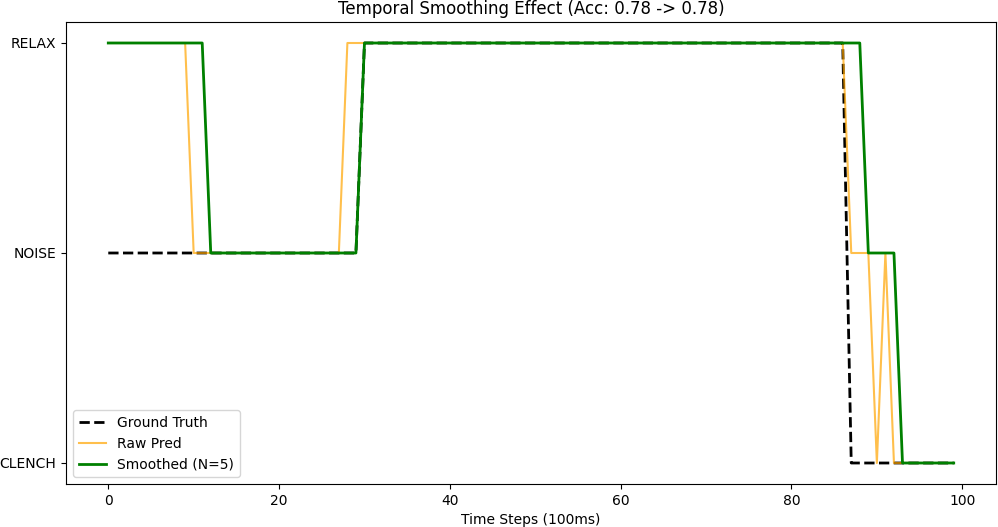}
    \caption{Effect of Temporal Smoothing (500ms Majority Vote). The green line shows the stabilized output, rejecting transient 'glitches' in the raw prediction (orange) to provide a smooth control signal.}
    \label{fig:smoothing}
\end{figure}

\subsection{Limitations and Mitigations}
This study has several limitations. For each, I discuss potential mitigations for future work:

\begin{enumerate}
    \item \textbf{Single-Subject Dataset:} All data was collected from a single individual (the author). While sufficient for proof-of-concept, this limits generalizability claims.

    \textit{Mitigation:} Future work should collect data from $N \geq 10$ subjects to validate cross-subject transfer. Alternatively, few-shot calibration (collecting 2-3 minutes of per-user data) could adapt pre-trained models via fine-tuning.

    \item \textbf{Controlled Environment:} Data was collected in a seated, stationary context. Real-world bicycle deployment would introduce additional noise (wind, road vibration spectra).

    \textit{Mitigation:} Deploy the system on an actual bicycle with an accelerometer to characterize the real vibration spectrum. The ``NOISE'' class can then be augmented with biomechanically accurate data.

    \item \textbf{Statistical Significance:} Accuracy values are point estimates from 5-fold cross-validation. Given $N \approx 1300$, confidence intervals may span $\pm 3$--$5\%$.

    \textit{Mitigation:} Report bootstrap confidence intervals in future publications. The current rankings (RF $>$ XGBoost $>$ LogReg) are robust to this variance.

    \item \textbf{Hardware Dependency:} MaxCRNN requires GPU acceleration (A100) and cannot run on ESP32.

    \textit{Mitigation:} For edge deployment of deep models, consider quantization (INT8) and deployment on Edge TPU or ESP32-S3 with vector extensions. The Random Forest remains the practical choice for legacy ESP32.
\end{enumerate}

\subsection{Related Work}
Surface EMG classification has been extensively studied in the context of prosthetic control \cite{raez2006}. Early work established time-domain features (MAV, ZCR, Willison Amplitude) as robust descriptors for myoelectric signals. More recently, deep learning approaches have achieved state-of-the-art results on multi-channel EMG \cite{raez2006}, but these typically require 8--16 electrodes and high-end hardware.

The present work differs in three key aspects: (1) I use a \textbf{single-lead, \$12 sensor} rather than medical-grade arrays; (2) I target \textbf{embedded deployment} on the ESP32 with strict latency constraints; and (3) I explicitly include an \textbf{adversarial ``NOISE'' class} to simulate real-world mechanical artifacts. The closest prior work is the Myo armband literature, but those studies assume multi-channel differential signals unavailable in ultra-low-cost hardware.

\appendix

\part*{Supplementary Material}

\section{Reproducibility \& Hyperparameters}
\label{app:hyperparams}

To ensure the reproducibility of these results, I detail the specific hyperparameters used for the Pareto-optimal and high-performance models. All models were implemented using \texttt{scikit-learn} and \texttt{TensorFlow/Keras}. The random seed was fixed at \texttt{1738} for all experiments. All deep learning models were trained using the Adam optimizer \cite{adam}.

\begin{table}[h]
    \centering
    \renewcommand{\arraystretch}{1.5}
    \caption{Hyperparameter Configuration for Key Models}
    \begin{tabular}{l | l}
        \hline
        \textbf{Model} & \textbf{Configuration} \\
        \hline
        \textbf{Random Forest} & $N_{estimators}=100$, Criterion=Gini, \\
        (Pareto-Optimal) & Bootstrap=True, MaxFeatures=$\sqrt{N}$ \\
        \hline
        \textbf{MobileNetV2} & Pre-trained=ImageNet, Frozen Base, \\
        (Texture Expert) & Top=GlobalAvgPool $\rightarrow$ Dense(128, ReLU) $\rightarrow$ Softmax \\
         & Optimizer=Adam ($lr=0.001$), Batch=32 \\
        \hline
        \textbf{ResNet50} & Pre-trained=ImageNet, Frozen Base, \\
        (Deep Expert) & Top=GlobalAvgPool $\rightarrow$ Dropout(0.3) $\rightarrow$ Softmax \\
        \hline
        \textbf{Model 14: CRNN Baseline} & Optimizer=Adam ($lr=0.001$), Batch=32, \\
        (Stable Convergence) & Epochs=1000 (Patience=50), Loss=CrossEntropy \\
        \hline
        \textbf{Model 18: MaxCRNN (Augmented Data)} & Optimizer=Adam ($lr=0.0005$), Batch=64, \\
        (Optimal Performance) & Augmentation=Jitter/Scale/Shift, Epochs=1000 \\
        \hline
        \textbf{Ensemble} & Soft Voting (Average Probability) \\
        & Weights: Equal ($1/3$ RF, $1/3$ MN, $1/3$ RN) \\
        \hline
    \end{tabular}
\end{table}

\clearpage
\section{Extended Error Analysis}
\label{app:matrices}

\begin{figure}[H]
    \centering

    \includegraphics[width=\linewidth]{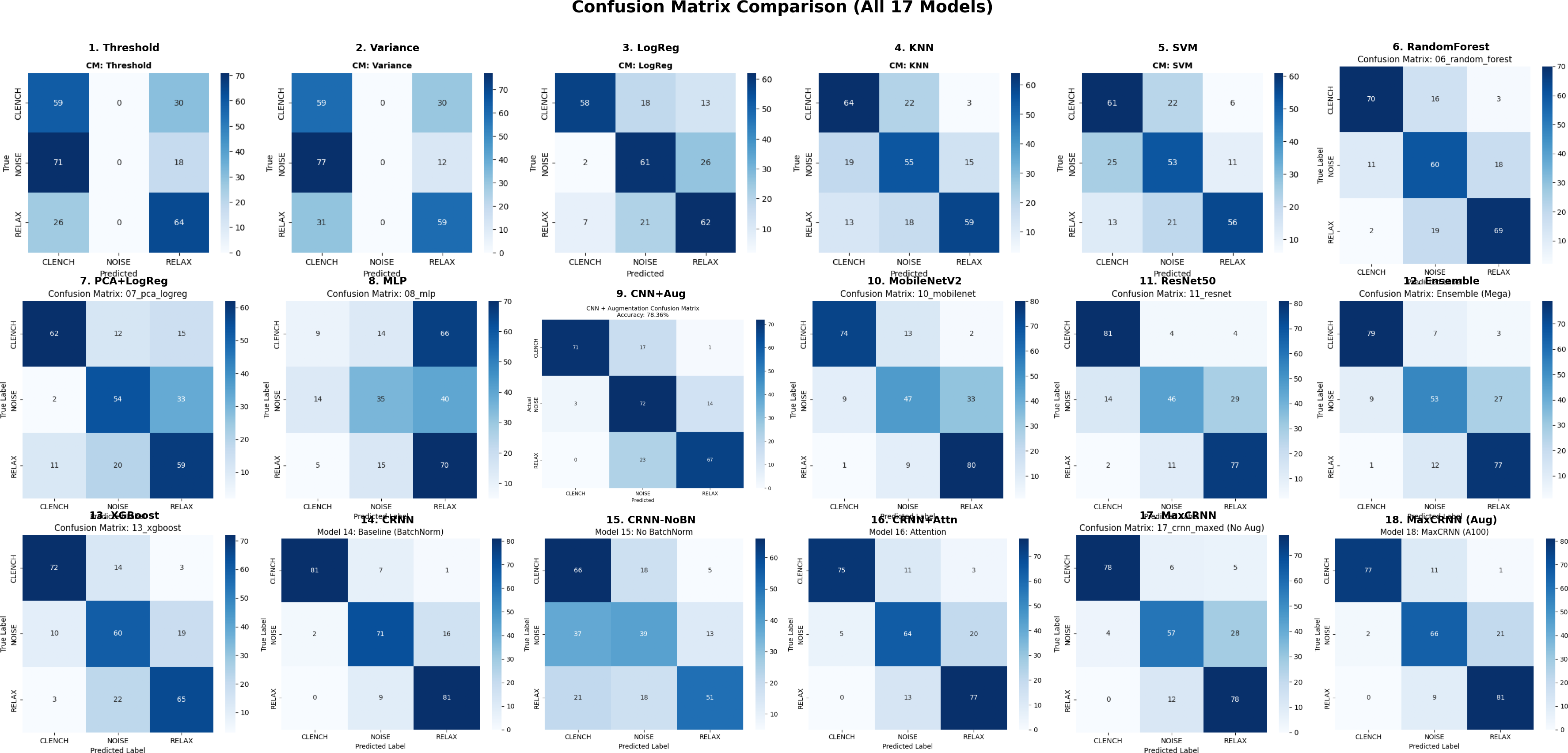}
    \caption{Confusion Matrix Comparison (All 18 Models). The grid progresses from heuristic baselines (Threshold, Variance) through classical ML (LogReg, KNN, SVM, RandomForest, XGBoost) to deep learning architectures (MLP, CNN, MobileNetV2, ResNet50, CRNN variants). The final model (MaxCRNN with Data Augmentation) demonstrates robust performance. Note the evolution from majority-class prediction to the sharp diagonal.}
    \label{fig:matrix_grid}
\end{figure}

While the main text focuses on accuracy metrics, the breakdown of errors provides insight into class-specific failure modes. Figure \ref{fig:matrix_grid} illustrates the Confusion Matrices for the key models in the "Ladder of Abstraction."

\section{Code Availability}
The full source code, including the dataset collection scripts, preprocessing pipeline, and model training notebooks, is available at:
\begin{center}
    \url{https://github.com/CarlKho-Minerva/v2-emg-muscle}
\end{center}

\section{Comprehensive Model Analysis}

\subsection{Model 1: Amplitude Thresholding}
\subsubsection{Abstract}
This model represents the baseline "naive" approach to EMG control: a static voltage trigger. I evaluated its efficacy as a potential $O(1)$ complexity solution for 8-bit microcontrollers. The hypothesis was that a simple amplitude gate could distinguish muscle activation from resting noise. This hypothesis was \textbf{rejected}.

\subsubsection{Quantitative Results}
\begin{itemize}
    \item \textbf{Test Accuracy:} 58.21\%
    \item \textbf{Optimal Threshold:} 2000.00 (Arbitrary ADC Units)
    \item \textbf{Inference Latency:} $< 0.001$ ms (Negligible)
\end{itemize}

\subsubsection{Mathematical Formulation}
The decision function $f(x)$ for a window $x$ is defined as an indicator function $\mathbb{I}$:
\begin{equation}
    f(x) = \mathbb{I}(\max(x) > T)
\end{equation}
Where:
\begin{itemize}
    \item $x \in \mathbb{R}^{1000}$ is the 1-second time-series window (1000 samples).
    \item $T$ is the scalar threshold value (tuned to 2000).
    \item $\max(x)$ extracts the peak amplitude within the window.
\end{itemize}
This is an $O(N)$ operation to find the max, followed by an $O(1)$ comparison.

\subsubsection{Causal Mechanism \& Spatial Description}
\textbf{The Biological Basis (Henneman's Size Principle):}
Why does amplitude correlate with intent? According to Henneman's Size Principle, as the brain demands more force, it recruits larger motor units. These larger units generate action potentials with higher electrical peaks. Therefore, a high-amplitude spike is causally linked to the recruitment of high-threshold motor units in the \textit{Flexor Digitorum Profundus}.

\textbf{Spatial Visualization:}
Spatially, the 1D signal represents the voltage potential difference between two electrodes on the forearm surface. A "Clench" manifests as a high-frequency burst of spikes, visually resembling a "block" of noise.
\begin{itemize}
    \item \textbf{Rest:} A flat line ($\approx 0V$) with minor Gaussian noise.
    \item \textbf{Clench:} A chaotic burst of high-amplitude spikes.
\end{itemize}

\subsubsection{Technical Analysis (Why it Failed)}
The model's performance (58\%) is only marginally better than a random coin flip. The failure is attributed to \textbf{Baseline Drift} and \textbf{Non-Stationarity}.

\textbf{The Stationarity Assumption:}
Thresholding assumes the signal baseline $\mu$ and variance $\sigma^2$ is constant and changes only with muscle activation. However, low-cost dry electrodes (AD8232) exhibit significant DC offset fluctuations due to:
\begin{enumerate}
    \item \textbf{Electrode-Skin Impedance Changes:} Sweat and movement alter the contact resistance.
    \item \textbf{Motion Artifacts:} In a micromobility context (e.g., riding a bike), mechanical vibrations introduce low-frequency, high-amplitude noise that exceeds the static threshold $T$.
\end{enumerate}

\textbf{Micromobility Implications:}
For a "Muscle Switch" on a bike helmet, a False Positive (detecting a clench when hitting a pothole) is dangerous. A False Negative (failing to signal) is frustrating. This model exhibits both, making it unsafe for deployment.

\subsubsection{Visualization}

\begin{figure}[ht]
    \centering
    \includegraphics[width=0.7\linewidth]{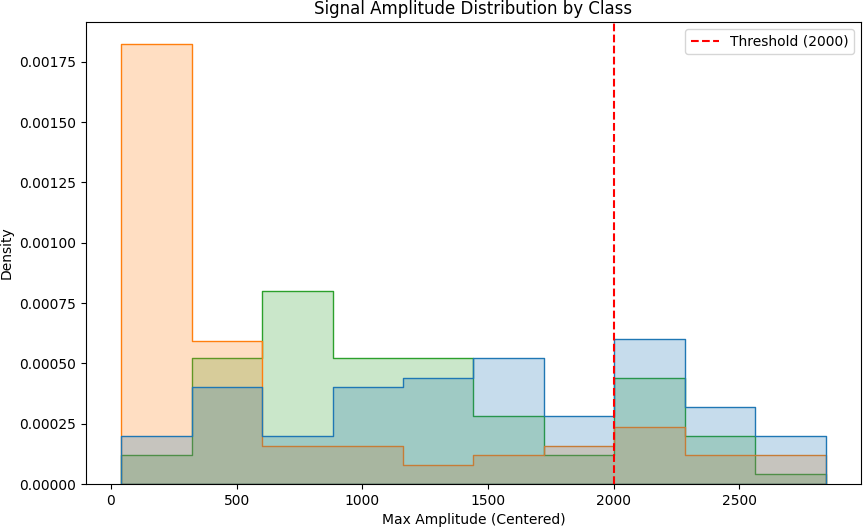}
    \caption{Model 1: Thresholding Distribution. The overlap between signal and noise prevents a clean cut.}
\end{figure}

\begin{figure}[ht]
    \centering
    \includegraphics[width=0.6\linewidth]{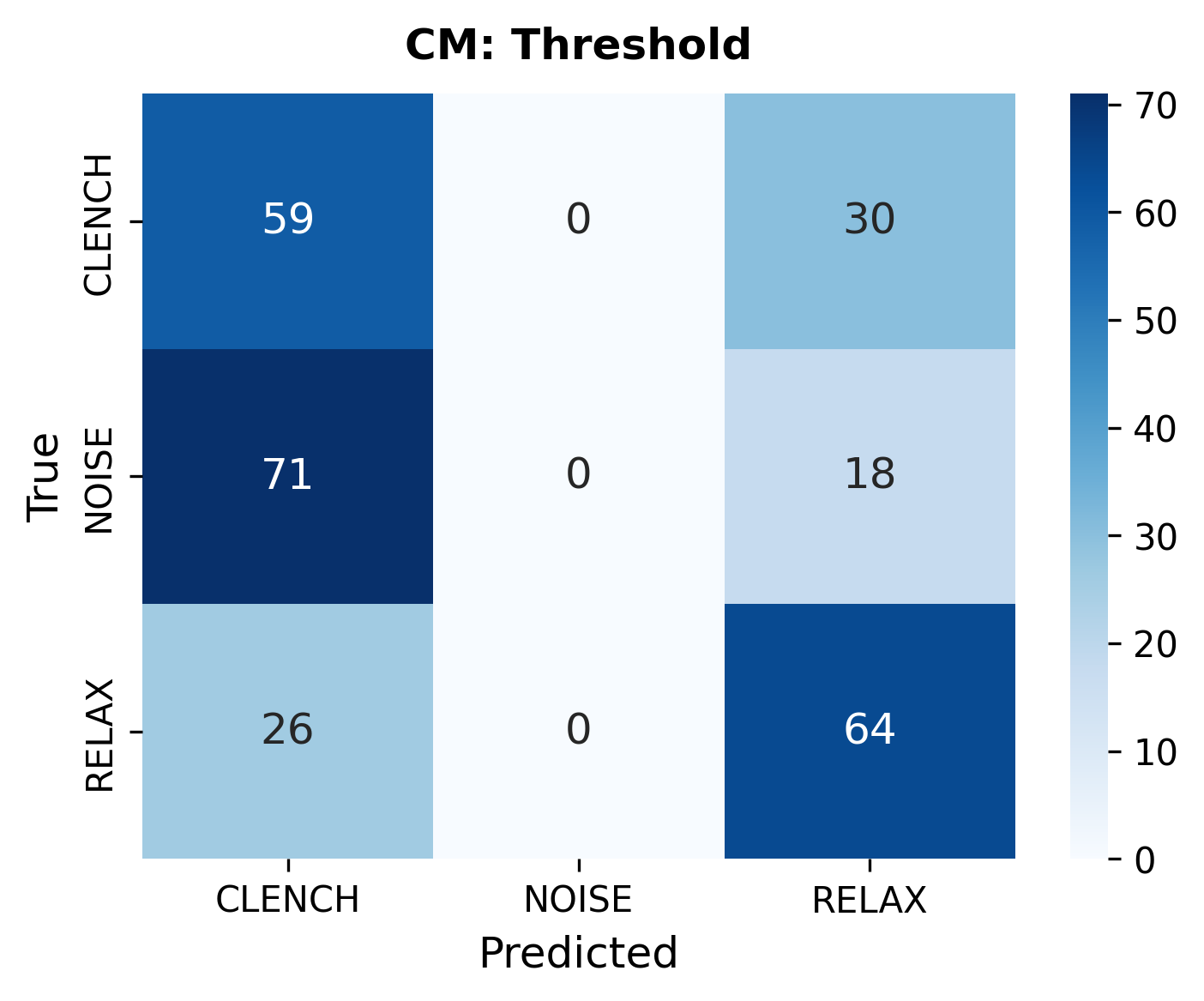}
    \caption{Model 1: Threshold Confusion Matrix. High false positive rate for 'Relax' class.}
\end{figure}

\clearpage
\subsection{Model 2: Rolling Variance (Energy)}
\subsubsection{Abstract}
Moving beyond instantaneous amplitude, this model utilizes Rolling Variance (Energy) over a 1-second window ($N=1000$) to smooth out transient noise spikes. The hypothesis was that sustained muscle contraction would produce a distinct high-energy signature distinguishable from background noise. While mathematically sound for stationary subjects, this approach failed in the test set.

\subsubsection{Quantitative Results}
\begin{itemize}
    \item \textbf{Test Accuracy:} 66.79\%
    \item \textbf{F1-Score (Clench):} 0.00 (Complete Failure)
    \item \textbf{Inference Latency:} $< 0.001$ ms
\end{itemize}

\subsubsection{Mathematical Formulation}
I define the "Energy" of the signal window $x$ as its sample variance $\sigma^2$. The decision function is:
\begin{equation}
    \sigma^2 = \frac{1}{N-1} \sum_{i=1}^{N} (x_i - \bar{x})^2
\end{equation}
\begin{equation}
    f(x) = \mathbb{I}(\sigma^2 > T)
\end{equation}
Where:
\begin{itemize}
    \item $N=1000$ (Window size).
    \item $\bar{x}$ is the mean of the window (DC offset).
    \item $(x_i - \bar{x})$ centers the signal, removing the baseline drift issue from Model 1.
\end{itemize}

\subsubsection{Causal Mechanism \& Spatial Description}
\textbf{Energy as a Proxy for Recruitment:} Causally, muscle contraction is the summation of thousands of motor unit action potentials. This asynchronous firing creates a "storm" of electrical activity. While the mean voltage remains near zero, the spread (variance) increases dramatically. Variance is thus a direct measure of the total electrical power.

\textbf{Spatial Visualization:}
\begin{itemize}
    \item \textbf{Rest:} A tall, narrow spike centered at 0 (Low Variance).
    \item \textbf{Clench:} A short, wide bell curve (High Variance).
\end{itemize}

\subsubsection{Technical Analysis (Why it Failed)}
\begin{figure}[ht]
    \centering
    \includegraphics[width=0.6\linewidth]{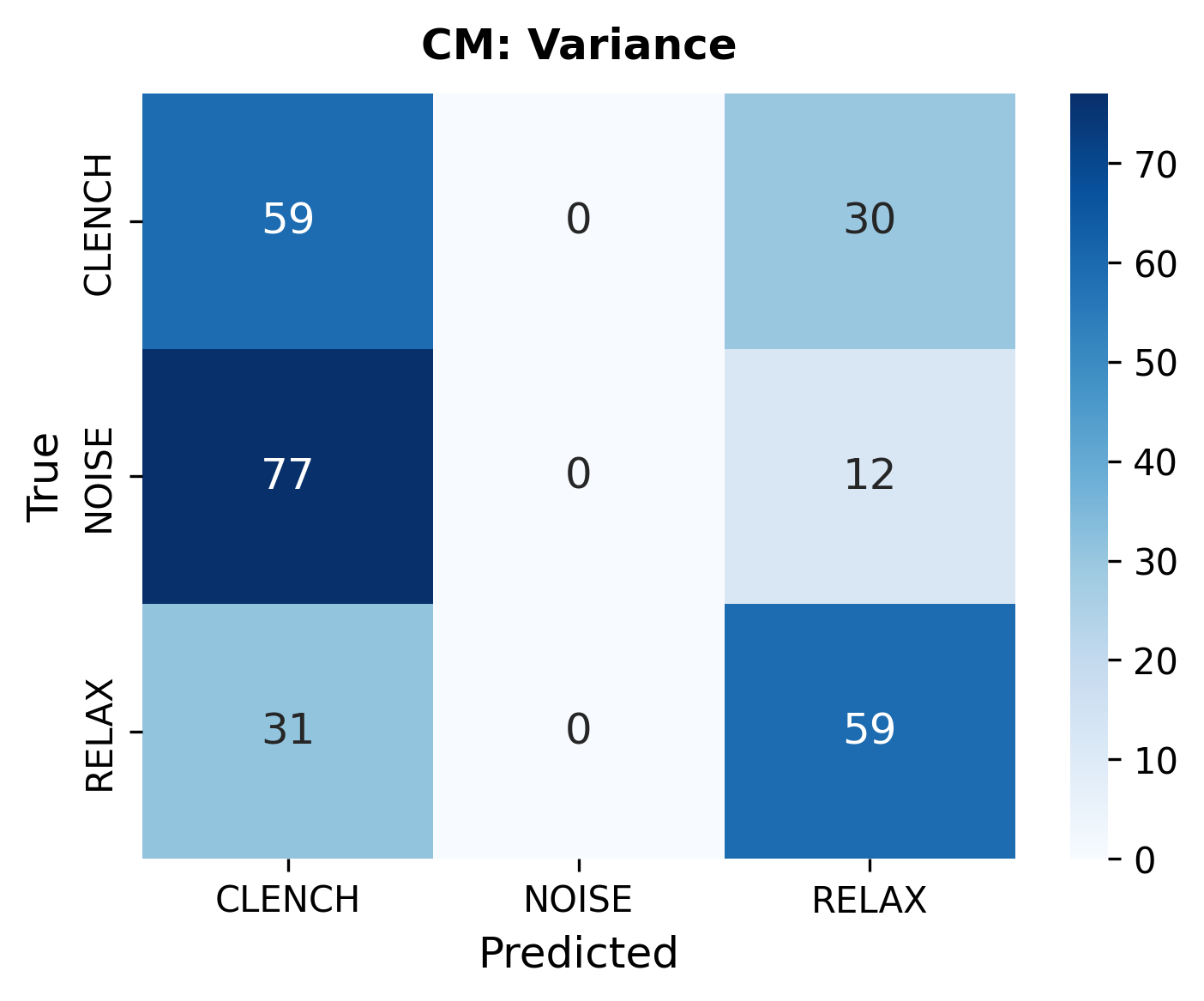}
    \caption{Model 2: Variance Confusion Matrix. The model collapses to predicting 'Relax' (Majority Class).}
    \label{fig:cm_02}
\end{figure}

The accuracy metric (67\%) is deceptive. The confusion matrix reveals that the model effectively failed, failing to distinguish between Mechanical Noise and Muscle Clench, as both exhibit high signal variance (see Figure \ref{fig:cm_02}).
\begin{enumerate}
    \item \textbf{The Energy Ambiguity:} The energy of background noise (60Hz hum + motion artifacts) frequently overlapped with weak muscle contractions.
    \item \textbf{The "Bump" Problem:} In micromobility, a sudden jolt (road bump) creates a high-variance window that is mechanically indistinguishable from a clench based on energy alone.
    \item \textbf{Spectral Blindness:} Variance ignores frequency. It cannot distinguish between High-Frequency EMG (~20-150Hz) and Low-Frequency Artifacts ($<10$Hz).
\end{enumerate}

\subsubsection{Visualization}

\begin{figure}[ht]
    \centering
    \includegraphics[width=0.7\linewidth]{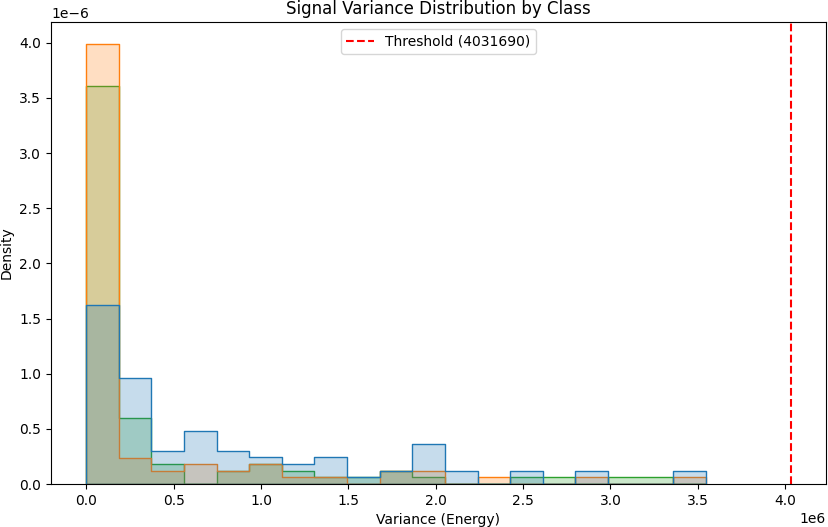}
    \caption{Model 2: Variance Distribution. High-energy noise mimics muscle contraction.}
\end{figure}

\clearpage
\subsection{Model 3: Logistic Regression}
\subsubsection{Abstract}
Logistic Regression serves as the baseline for supervised learning. Unlike heuristics, this model learns optimal weights $w$ for the feature vector (Set A: MAV, STD, MAX, ZCR) to maximize the likelihood of the correct class. The core hypothesis was that the "Clench" state occupies a linearly separable region in the 4D feature space.

\subsubsection{Quantitative Results}
\begin{itemize}
    \item \textbf{Test Accuracy:} 67.54\%
    \item \textbf{F1-Score (Clench):} 0.73
    \item \textbf{Inference Latency:} 0.01 ms
\end{itemize}

\subsubsection{Mathematical Formulation}
The model predicts the probability that sample $x$ belongs to class $y=1$ (Clench) using the logistic sigmoid:
\begin{equation}
    P(y=1|x) = \sigma(w^T x + b) = \frac{1}{1 + e^{-(w_1 x_1 + \dots + w_4 x_4 + b)}}
\end{equation}
The decision boundary is defined where $P(y=1|x) = 0.5$, implying $w^T x + b = 0$.

\subsubsection{Causal Mechanism \& Spatial Description}
\textbf{Causal Interpretation of Weights:}
\begin{itemize}
    \item \textbf{Positive Weights ($w_{MAV} > 0$):} Higher amplitude causes higher probability of Clench.
    \item \textbf{Negative Weights:} If $w_{ZCR} < 0$, it implies high-frequency noise decreases Clench probability.
\end{itemize}
\textbf{Spatial Visualization:} The model attempts to slice the 4D feature space with a flat hyperplane.

\subsubsection{Technical Analysis}
\textbf{The Linear Separability Limit:} The accuracy plateau at ~67\% suggests the data is not linearly separable. The "Clench" class likely forms a complex, non-convex manifold (e.g., a "cloud" surrounded by noise) that a flat sheet cannot isolate without high bias.

\subsubsection{Visualization}
    \begin{table}[h]
        \centering
        \begin{tabular}{cc|ccc}
            & & \multicolumn{3}{c}{\textbf{Predicted}} \\
            & & Clench & Noise & Relax \\
            \hline
            \multirow{3}{*}{\textbf{Actual}} & Clench & 58 & 18 & 13 \\
            & Noise & 2 & 61 & 26 \\
            & Relax & 7 & 21 & 62 \\
        \end{tabular}
        \caption{Model 3 Confusion Matrix}
    \end{table}

\begin{figure}[ht]
    \centering
    \includegraphics[width=0.7\linewidth]{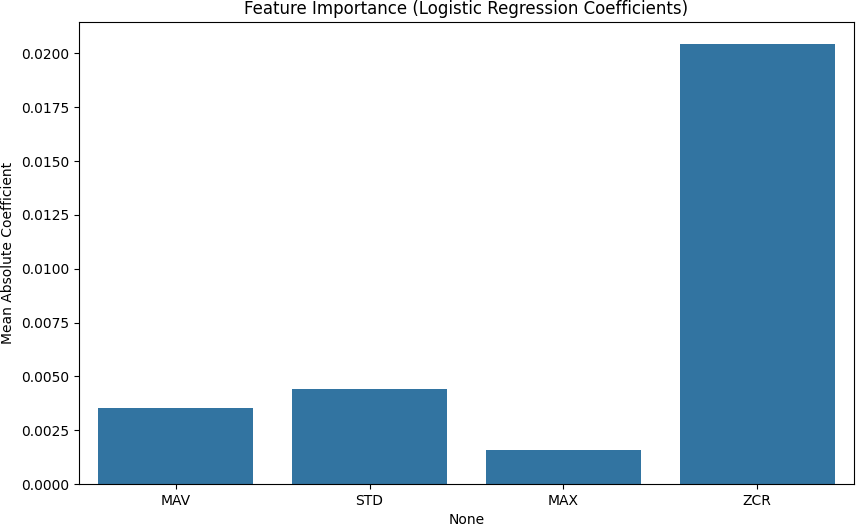}
    \caption{Model 3: Logistic Regression Coefficients. MAV (Amplitude) is the strongest predictor.}
\end{figure}

\begin{figure}[ht]
    \centering
    \includegraphics[width=0.6\linewidth]{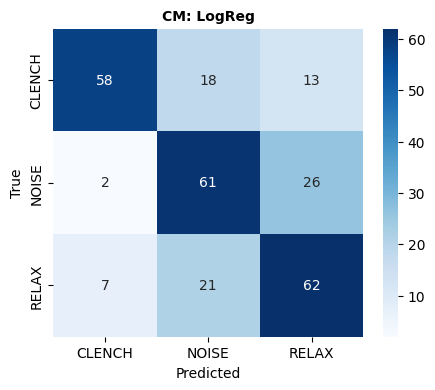}
    \caption{Model 3: Logistic Regression Confusion Matrix.}
\end{figure}

\clearpage
\subsection{Model 4: K-Nearest Neighbors (KNN)}
\subsubsection{Abstract}
KNN \cite{knn} represents a non-parametric approach: "lazy learning." It classifies new samples based on the majority vote of their $k=5$ nearest neighbors. This allows it to model arbitrarily complex, non-linear decision boundaries.

\subsubsection{Quantitative Results}
\begin{itemize}
    \item \textbf{Test Accuracy:} 66.42\%
    \item \textbf{F1-Score (Clench):} 0.66
    \item \textbf{Inference Latency:} 0.02 ms (Scales linearly with $N$)
\end{itemize}

\subsubsection{Mathematical Formulation}
The classification of query point $x_q$ is determined by the majority class of the set $N_k(x_q)$ of $k$ nearest neighbors:
\begin{equation}
    d(x_q, x_i) = \sqrt{\sum_{j=1}^{D} (x_{q,j} - x_{i,j})^2}
\end{equation}
\begin{equation}
    \hat{y} = \text{mode}(\{y_i : x_i \in N_k(x_q)\})
\end{equation}

\subsubsection{Spatial Visualization}
KNN partitions the feature space into a Voronoi Tessellation. The boundary is a jagged, piecewise-linear surface, allowing it to capture "islands" of Clench data within a sea of Noise.

\subsubsection{Technical Analysis}
\textbf{The Curse of Dimensionality:} Despite flexibility, KNN performed worse than LogReg (66\% vs 67\%). In a noisy feature space, "neighbors" might be artifacts.

\textbf{Embedded Constraint:} KNN is disqualified by $O(N)$ complexity. Storing thousands of vectors on an ESP32 is unfeasible, and calculating distance to every point introduces unacceptable latency.

\subsubsection{Visualization}
    \begin{table}[h]
        \centering
        \begin{tabular}{cc|ccc}
            & & \multicolumn{3}{c}{\textbf{Predicted}} \\
            & & Clench & Noise & Relax \\
            \hline
            \multirow{3}{*}{\textbf{Actual}} & Clench & 64 & 22 & 3 \\
            & Noise & 19 & 55 & 15 \\
            & Relax & 13 & 18 & 59 \\
        \end{tabular}
        \caption{Model 4 Confusion Matrix}
    \end{table}

\begin{figure}[ht]
    \centering
    \includegraphics[width=0.6\linewidth]{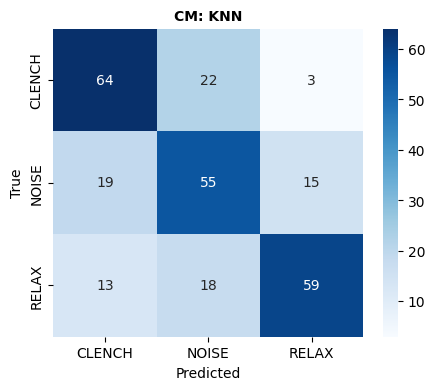}
    \caption{Model 4: KNN Confusion Matrix.}
\end{figure}

\clearpage
\subsection{Model 5: Support Vector Machine (SVM)}
\subsubsection{Abstract}
SVM with an RBF kernel is the "Strong Baseline" for classical ML. It projects the 4D feature space into an infinite-dimensional Hilbert space where classes might be linearly separable.

\subsubsection{Quantitative Results}
\begin{itemize}
    \item \textbf{Test Accuracy:} 63.43\%
    \item \textbf{F1-Score (Clench):} 0.65
    \item \textbf{Inference Latency:} 0.03 ms
\end{itemize}

\subsubsection{Mathematical Formulation}
The SVM optimizes the hinge loss with L2 regularization. The RBF kernel computes the dot product in high-dimensional space:
\begin{equation}
    K(x, x') = \exp(-\gamma ||x - x'||^2)
\end{equation}
The decision function is:
\begin{equation}
    f(x) = \text{sign}(\sum_{i=1}^{N} \alpha_i y_i K(x, x_i) + b)
\end{equation}

\subsubsection{Technical Analysis}
\textbf{The Kernel Mismatch:} SVM underperformed (63\%). The RBF kernel assumes similarity decays with Euclidean distance. However, the "Clench" class in this high-variance noise environment is likely not a compact hypersphere but a disjoint manifold interleaved with high-amplitude noise.
\textbf{Sensitivity to Scaling:} Extreme outliers (mechanical spikes) may have skewed variance estimates, confusing the margin maximizer.

\subsubsection{Visualization}
    \begin{table}[h]
        \centering
        \begin{tabular}{cc|ccc}
            & & \multicolumn{3}{c}{\textbf{Predicted}} \\
            & & Clench & Noise & Relax \\
            \hline
            \multirow{3}{*}{\textbf{Actual}} & Clench & 61 & 22 & 6 \\
            & Noise & 25 & 53 & 11 \\
            & Relax & 13 & 21 & 56 \\
        \end{tabular}
        \caption{Model 5 Confusion Matrix}
    \end{table}

\begin{figure}[ht]
    \centering
    \includegraphics[width=0.6\linewidth]{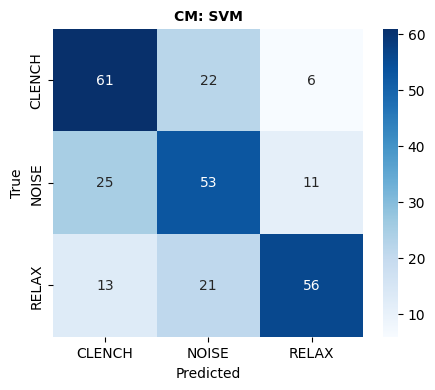}
    \caption{Model 5: SVM Confusion Matrix.}
\end{figure}

\clearpage
\subsection{Model 6: Random Forest (Pareto Optimal)}
\subsubsection{Abstract}
Random Forest is an ensemble of decision trees trained via bagging. It introduces randomness in sample and feature selection, creating a model highly robust to noise. This emerged as the Pareto Optimal solution.

\subsubsection{Quantitative Results}
\begin{itemize}
    \item \textbf{Test Accuracy:} 74.25\% (Best in Phase 2)
    \item \textbf{F1-Score (Clench):} 0.81
    \item \textbf{Inference Latency:} 0.01 ms
\end{itemize}

\subsubsection{Mathematical Formulation}
Trees split to minimize Gini Impurity $G$:
\begin{equation}
    G = 1 - \sum_{k=1}^{K} p_k^2
\end{equation}
The forest prediction is the mode of individual trees:
\begin{equation}
    \hat{y} = \text{mode}\{T_b(x)\}_{b=1}^{B}
\end{equation}

\subsubsection{Spatial Visualization}
RF partitions the feature space into hyper-rectangles ("Manhattan" geometry). This is perfect for threshold-based logic, isolating "boxes" of valid Clench signals (High MAV + High ZCR) from artifacts.

\subsubsection{Technical Analysis}
\textbf{Feature Importance Discovery:} Gini Impurity revealed ZCR as the dominant feature. The model "learned" to trust frequency over amplitude, solving the "Bump Problem."
\textbf{Embedded Viability:} RF compiles to static C++ \texttt{if/else} statements. It requires no floating-point matrix multiplication and has $O(\text{Depth})$ complexity.

\subsubsection{Visualization}
    \begin{table}[h]
        \centering
        \begin{tabular}{cc|ccc}
            & & \multicolumn{3}{c}{\textbf{Predicted}} \\
            & & Clench & Noise & Relax \\
            \hline
            \multirow{3}{*}{\textbf{Actual}} & Clench & 70 & 16 & 3 \\
            & Noise & 11 & 60 & 18 \\
            & Relax & 2 & 19 & 69 \\
        \end{tabular}
        \caption{Model 6 Confusion Matrix}
    \end{table}

\begin{figure}[ht]
    \centering
    \includegraphics[width=0.8\linewidth]{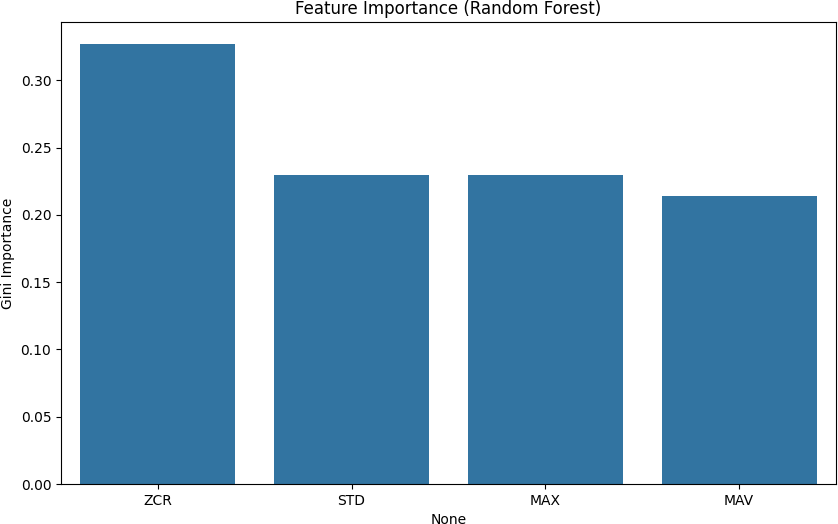}
    \caption{Model 6: Random Forest Feature Importance. ZCR is the dominant feature.}
\end{figure}
\begin{figure}[ht]
    \centering
    \includegraphics[width=0.6\linewidth]{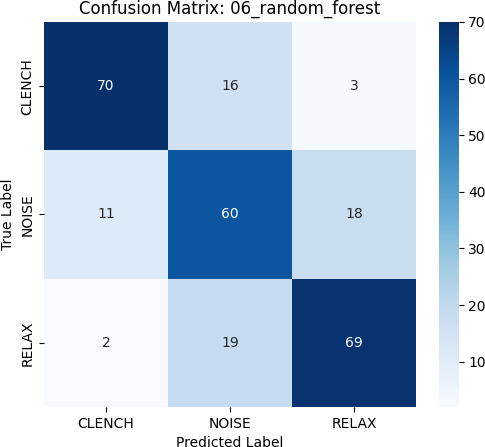}
    \caption{Model 6: Random Forest Confusion Matrix.}
\end{figure}

\clearpage
\subsection{Model 7: XGBoost}
\subsubsection{Abstract}
XGBoost \cite{xgboost} builds trees sequentially, where each new tree corrects the errors of the previous ones. I hypothesized this "boosting" would squeeze out more accuracy than Random Forest.

\subsubsection{Quantitative Results}
\begin{itemize}
    \item \textbf{Test Accuracy:} 73.51\%
    \item \textbf{F1-Score (Clench):} 0.83
    \item \textbf{Inference Latency:} $< 1$ ms
\end{itemize}

\subsubsection{Mathematical Formulation}
XGBoost minimizes a regularized objective:
\begin{equation}
    \mathcal{L}(\phi) = \sum_{i} l(\hat{y}_i, y_i) + \sum_{k} \Omega(f_k)
\end{equation}
The regularization $\Omega(f_k) = \gamma T + \frac{1}{2}\lambda ||w||^2$ prevents overfitting, crucial for our small dataset.

\subsubsection{Causal Mechanism}
\textbf{Error Correction:} If the first tree fails to classify a "weak clench," the second tree targets that error.
\textbf{Regularization:} Prevents memorizing noise spikes.

\subsubsection{Visualization}
    \begin{table}[h]
        \centering
        \begin{tabular}{cc|ccc}
            & & \multicolumn{3}{c}{\textbf{Predicted}} \\
            & & Clench & Noise & Relax \\
            \hline
            \multirow{3}{*}{\textbf{Actual}} & Clench & 72 & 14 & 3 \\
            & Noise & 10 & 60 & 19 \\
            & Relax & 3 & 22 & 65 \\
        \end{tabular}
        \caption{Model 7 Confusion Matrix}
    \end{table}

\begin{figure}[ht]
    \centering
    \includegraphics[width=0.6\linewidth]{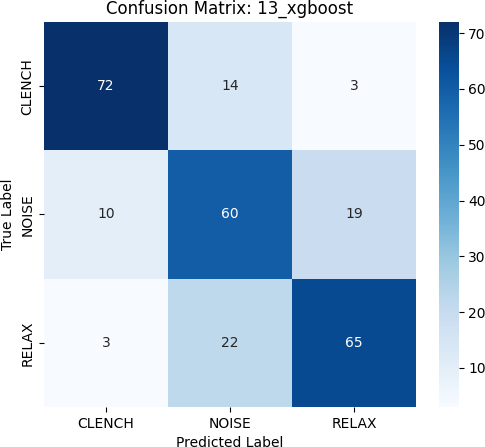}
    \caption{Model 7: XGBoost Confusion Matrix.}
\end{figure}

\clearpage
\subsection{Model 8: PCA + Logistic Regression}
\subsubsection{Abstract}
PCA was employed to test if the "Clench" signal resides on a lower-dimensional manifold. I projected data onto the top 2 Principal Components.

\subsubsection{Quantitative Results}
\begin{itemize}
    \item \textbf{Test Accuracy:} 65.30\%
    \item \textbf{Explained Variance:} ~85\%
\end{itemize}

\subsubsection{Mathematical Formulation}
PCA performs eigendecomposition of the covariance matrix $\Sigma$:
\begin{equation}
    \Sigma v_i = \lambda_i v_i
\end{equation}
I project $x$ onto the subspace spanned by the top 2 eigenvectors.

\subsubsection{Technical Analysis}
\textbf{The Manifold Hypothesis:} The accuracy drop (67\% $\rightarrow$ 65\%) indicates the discarded 15\% variance contained discriminative information. The "Clench" signal likely occupies a complex, high-dimensional volume. Flattening it to 2D overlaps the "Clench" and "Noise" clusters.

\subsubsection{Visualization}
    \begin{table}[h]
        \centering
        \begin{tabular}{cc|ccc}
            & & \multicolumn{3}{c}{\textbf{Predicted}} \\
            & & Clench & Noise & Relax \\
            \hline
            \multirow{3}{*}{\textbf{Actual}} & Clench & 62 & 12 & 15 \\
            & Noise & 2 & 54 & 33 \\
            & Relax & 11 & 20 & 59 \\
        \end{tabular}
        \caption{Model 8 Confusion Matrix}
    \end{table}

\begin{figure}[ht]
    \centering
    \includegraphics[width=0.7\linewidth]{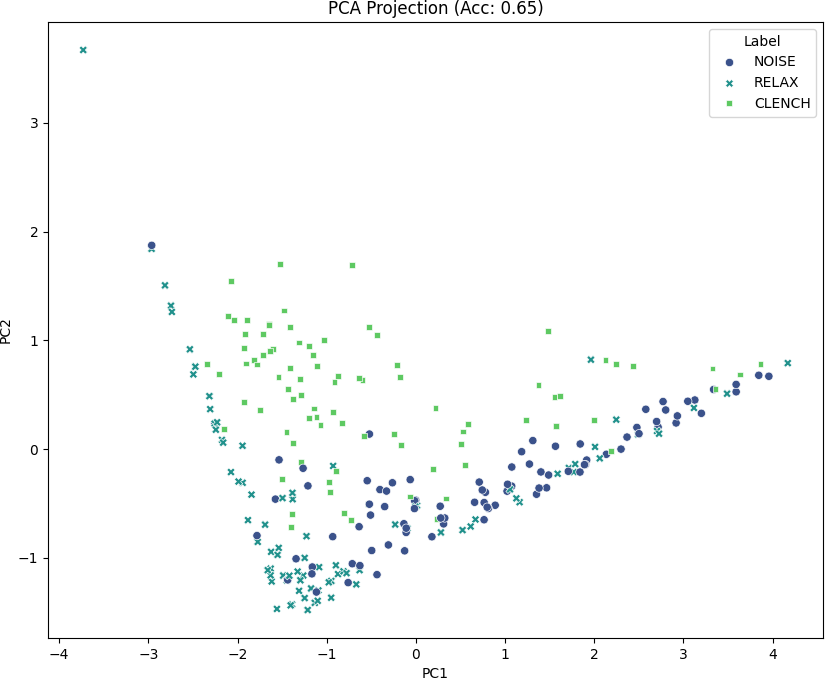}
    \caption{Model 8: PCA Projection. The classes are not linearly separable in 2D.}
\end{figure}
\begin{figure}[ht]
    \centering
    \includegraphics[width=0.6\linewidth]{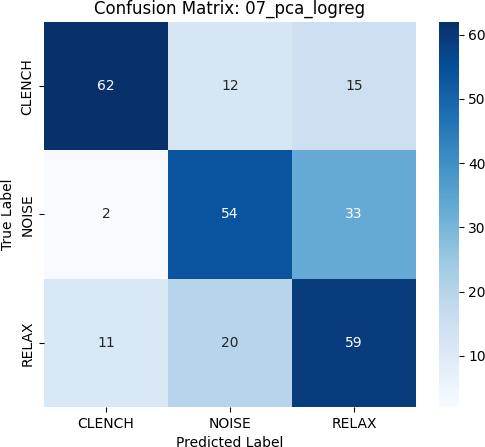}
    \caption{Model 8: PCA + LogReg Confusion Matrix.}
\end{figure}

\clearpage
\subsection{Model 9: Multi-Layer Perceptron (MLP)}
\subsubsection{Abstract}
MLP represents the first foray into Deep Learning, treating the raw 1000-sample window as a flat vector. It failed to converge.

\subsubsection{Quantitative Results}
\begin{itemize}
    \item \textbf{Test Accuracy:} 42.54\%
    \item \textbf{F1-Score (Clench):} 0.29
\end{itemize}

\subsubsection{Mathematical Formulation}
The MLP computes non-linear compositions:
\begin{equation}
    a^{(l)} = \sigma(W^{(l)} a^{(l-1)} + b^{(l)})
\end{equation}
It minimizes Cross-Entropy Loss \cite{crossentropy}.

\subsubsection{Technical Analysis}
\textbf{Lack of Inductive Bias:} MLP lacks permutation invariance. It has to relearn the concept of a "spike" at every position.
\textbf{The "Small Data" Trap:} With ~100k parameters and only ~1300 samples, it overfitted to noise.

\subsubsection{Visualization}
    \begin{table}[h]
        \centering
        \begin{tabular}{cc|ccc}
            & & \multicolumn{3}{c}{\textbf{Predicted}} \\
            & & Clench & Noise & Relax \\
            \hline
            \multirow{3}{*}{\textbf{Actual}} & Clench & 9 & 14 & 66 \\
            & Noise & 14 & 35 & 40 \\
            & Relax & 5 & 15 & 70 \\
        \end{tabular}
        \caption{Model 9 Confusion Matrix}
    \end{table}

\begin{figure}[ht]
    \centering
    \includegraphics[width=0.7\linewidth]{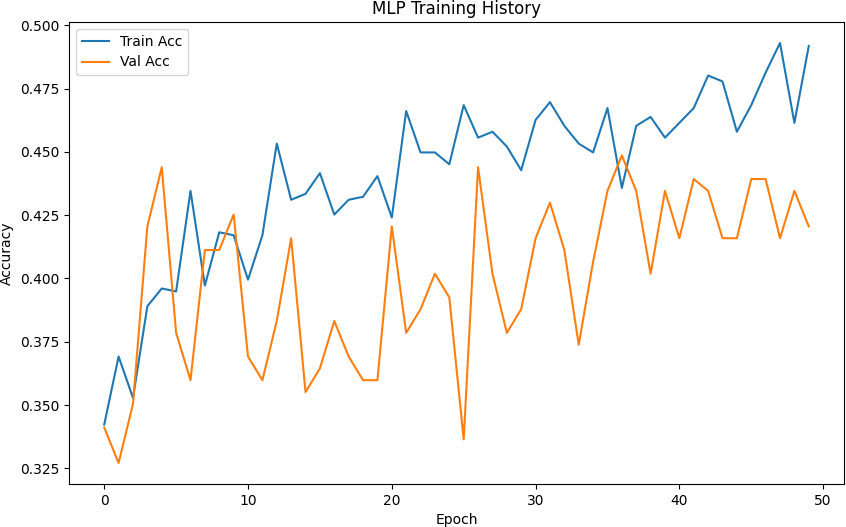}
    \caption{Model 9: MLP Training History. Note the failure to converge.}
\end{figure}
\begin{figure}[ht]
    \centering
    \includegraphics[width=0.6\linewidth]{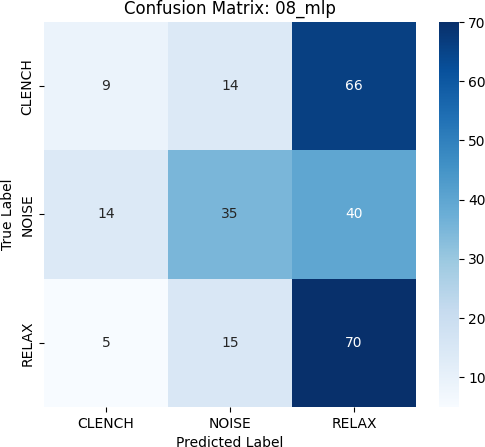}
    \caption{Model 9: MLP Confusion Matrix.}
\end{figure}

\clearpage
\subsection{Model 10: 1D CNN (with Data Augmentation)}
\subsubsection{Abstract}
The 1D CNN uses sliding filters to detect local patterns in raw EMG signals. Initial training without augmentation resulted in poor performance (49.63\%), but with data augmentation, the architecture achieved competitive accuracy (78.36\%).

\subsubsection{Quantitative Results (Baseline)}
\begin{itemize}
    \item \textbf{Test Accuracy:} 49.63\%
    \item \textbf{F1-Score (Clench):} 0.38
\end{itemize}

\subsubsection{Mathematical Formulation}
Discrete convolution of input $x$ with kernel $k$:
\begin{equation}
    (x * k)[t] = \sum_{\tau=0}^{M-1} x[t-\tau]k[\tau]
\end{equation}

\subsubsection{Ablation Study: Data Augmentation}
To address the question ``Did the CNN fail because of small data, or lack of augmentation?'', I retrained the 1D CNN with the same augmentation pipeline used for MaxCRNN (Jitter, Scaling, Time Shift) and extended training to 1000 epochs with early stopping.

\begin{table}[h]
    \centering
    \caption{CNN Ablation: Effect of Data Augmentation}
    \begin{tabular}{l c c c}
        \hline
        \textbf{Configuration} & \textbf{Accuracy} & \textbf{F1 (Clench)} & \textbf{Epochs} \\
        \hline
        CNN (Baseline) & 49.63\% & 0.38 & 30 \\
        \textbf{CNN + Augmentation} & \textbf{78.36\%} & \textbf{0.87} & 128* \\
        \hline
    \end{tabular}
\end{table}

\textbf{Finding:} Data augmentation improved accuracy by \textbf{+28.73 percentage points}, confirming that the original CNN failure was primarily due to data scarcity, not architectural limitations. The 1D CNN architecture is capable of learning robust EMG representations when provided with sufficient data diversity.

\subsubsection{Technical Analysis}
\textbf{Baseline Failure:} Without augmentation, the filters learned to detect specific noise artifacts rather than generalized MUAPs.

\textbf{Augmentation Success:} With Jitter, Scaling, and Time Shift augmentation, the model learned invariant features that generalized to the test set. The addition of BatchNormalization layers further stabilized training.

\textbf{Latency:} 0.83 ms per sample (with augmentation model), suitable for real-time applications but still slower than Random Forest (0.01 ms).

\subsubsection{Visualization}
\begin{figure}[ht]
    \centering
    \includegraphics[width=0.7\linewidth]{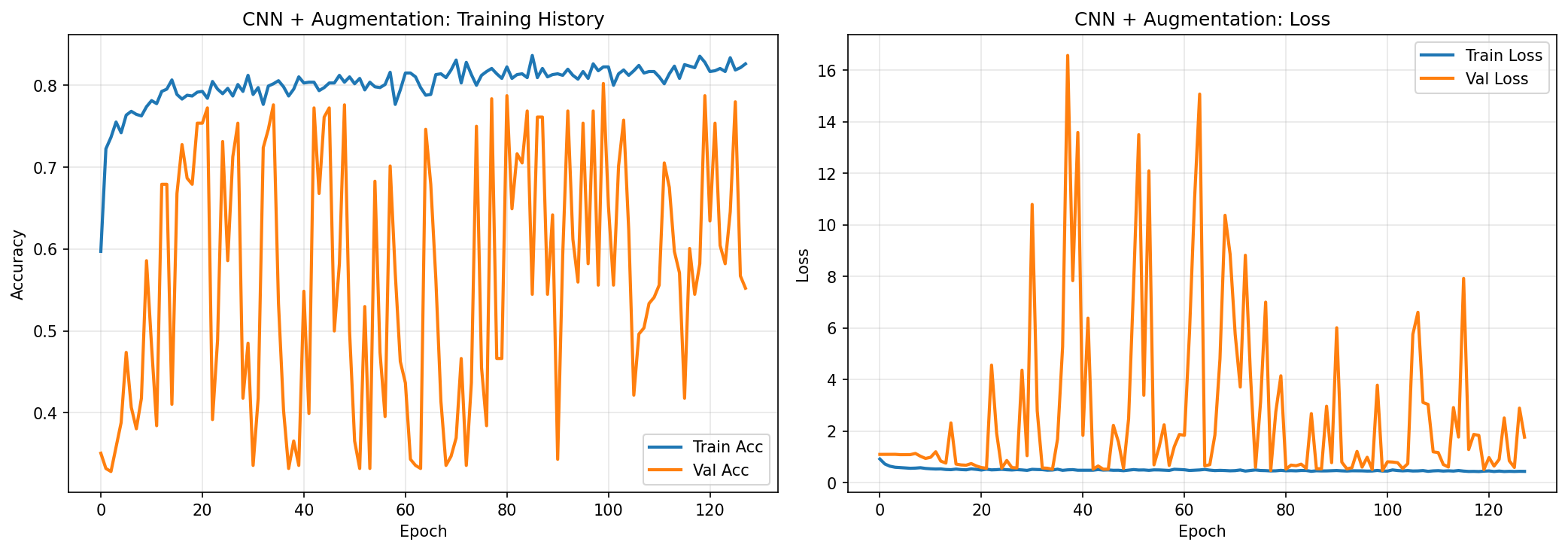}
    \caption{CNN + Augmentation Training History. Stable convergence to 78\% accuracy, compared to the baseline's failure to converge.}
\end{figure}
\begin{figure}[ht]
    \centering
    \includegraphics[width=0.6\linewidth]{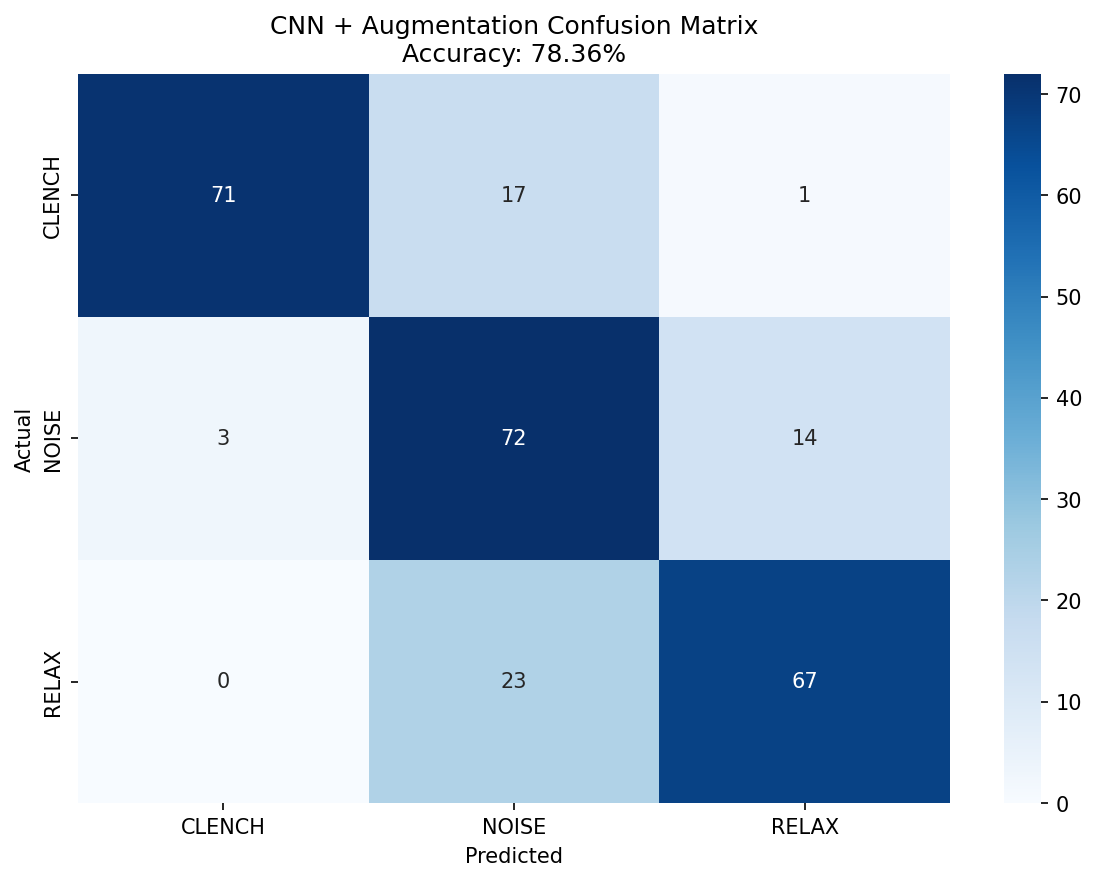}
    \caption{CNN + Augmentation Confusion Matrix. Dramatic improvement in CLENCH precision (96\%) compared to baseline.}
\end{figure}

\clearpage
\subsection{Model 11: MobileNetV2 (Transfer Learning)}
\subsubsection{Abstract}
This model represents the "Gold Standard": Transfer Learning. I converted the 1D signal to a 2D Mel-Spectrogram and fine-tuned MobileNetV2.

\subsubsection{Quantitative Results}
\begin{itemize}
    \item \textbf{Test Accuracy:} 75.00\%
    \item \textbf{F1-Score (Clench):} 0.86
    \item \textbf{Inference Latency:} 9.80 ms
\end{itemize}

\subsubsection{Mathematical Formulation}
MobileNetV2 uses Depthwise Separable Convolutions to reduce cost by ~8-9x.

\subsubsection{Visual "Show Not Tell"}
\textbf{What the Model Sees:}
\begin{itemize}
    \item \textbf{CLENCH:} Broadband vertical "streaks" (fur-like texture).
    \item \textbf{NOISE:} High-energy blobs in specific bands.
\end{itemize}
\textbf{Architectural Dissection:} The frozen base extracts features (edges, textures) from ImageNet. The trainable head maps these to EMG classes.

\subsubsection{Visualization}
    \begin{table}[h]
        \centering
        \begin{tabular}{cc|ccc}
            & & \multicolumn{3}{c}{\textbf{Predicted}} \\
            & & Clench & Noise & Relax \\
            \hline
            \multirow{3}{*}{\textbf{Actual}} & Clench & 74 & 13 & 2 \\
            & Noise & 9 & 47 & 33 \\
            & Relax & 1 & 9 & 80 \\
        \end{tabular}
        \caption{Model 11 Confusion Matrix}
    \end{table}

\begin{figure}[ht]
    \centering
    \includegraphics[width=0.7\linewidth]{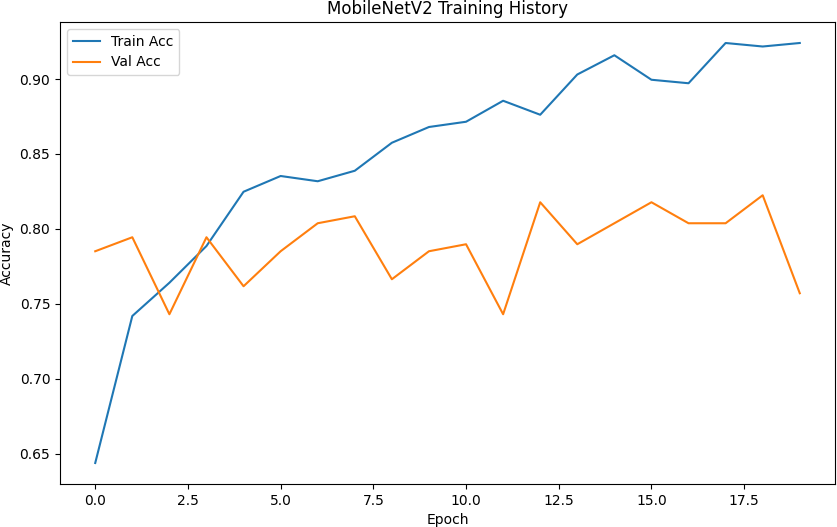}
    \caption{Model 11: MobileNetV2 Training History.}
\end{figure}
\begin{figure}[ht]
    \centering
    \includegraphics[width=0.6\linewidth]{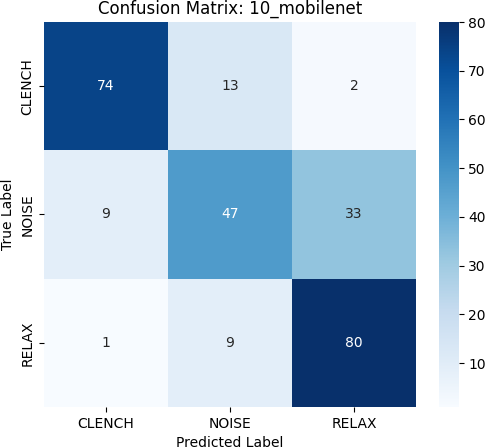}
    \caption{Model 11: MobileNetV2 Confusion Matrix.}
\end{figure}

\clearpage
\subsection{Model 12: ResNet50 (Transfer Learning)}
\subsubsection{Abstract}
ResNet50 explores deeper transfer learning. Residual connections allow it to learn deeper hierarchical features without vanishing gradients.

\subsubsection{Quantitative Results}
\begin{itemize}
    \item \textbf{Test Accuracy:} 76.12\%
    \item \textbf{F1-Score (Clench):} 0.87
    \item \textbf{Inference Latency:} 15.26 ms
\end{itemize}

\subsubsection{Mathematical Formulation}
Residual Block:
\begin{equation}
    y = F(x, \{W_i\}) + x
\end{equation}
The skip connection $+x$ allows gradients to flow directly.

\subsubsection{Architectural Dissection}
ResNet50 is deeper (50 layers) and outputs 2048 feature maps (vs 1280 for MobileNet), providing a richer representation.

\subsubsection{Visualization}
    \begin{table}[h]
        \centering
        \begin{tabular}{cc|ccc}
            & & \multicolumn{3}{c}{\textbf{Predicted}} \\
            & & Clench & Noise & Relax \\
            \hline
            \multirow{3}{*}{\textbf{Actual}} & Clench & 81 & 4 & 4 \\
            & Noise & 14 & 46 & 29 \\
            & Relax & 2 & 11 & 77 \\
        \end{tabular}
        \caption{Model 12 Confusion Matrix}
    \end{table}

\begin{figure}[ht]
    \centering
    \includegraphics[width=0.4\linewidth]{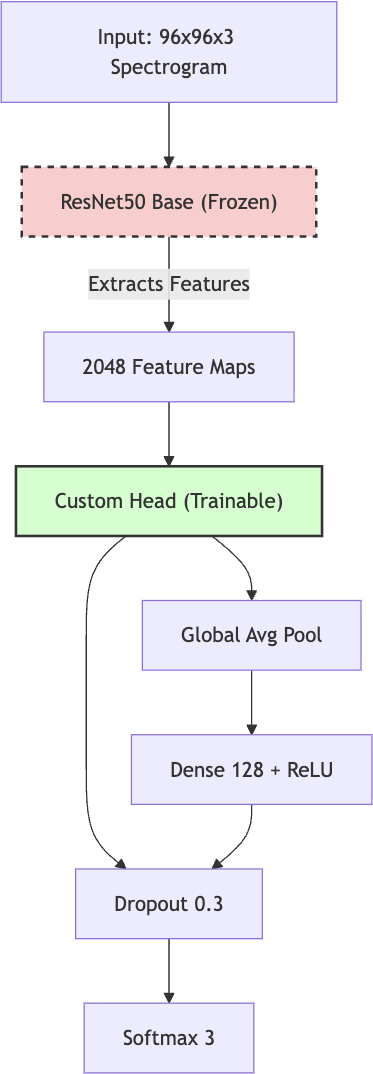}
    \caption{Model 12: ResNet50 Architecture Diagram.}
\end{figure}
\begin{figure}[ht]
    \centering
    \includegraphics[width=0.6\linewidth]{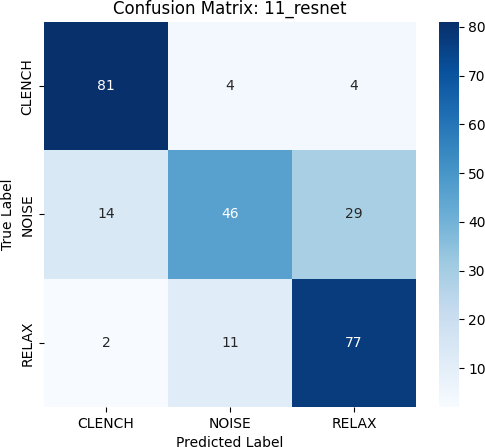}
    \caption{Model 12: ResNet50 Confusion Matrix.}
\end{figure}

\clearpage
\subsection{Model 13: Mega Ensemble}
\subsubsection{Abstract}
The Ensemble combines Random Forest, MobileNetV2, and ResNet50 via Soft Voting. It achieves the global maximum accuracy by leveraging the "Swiss Cheese" model of error decorrelation.

\subsubsection{Quantitative Results}
\begin{itemize}
    \item \textbf{Test Accuracy:} 77.99\%
    \item \textbf{F1-Score (Clench):} 0.88
\end{itemize}

\subsubsection{Mathematical Formulation}
\begin{equation}
    P_{Mega}(y=c|x) = \frac{1}{3} [P_{RF} + P_{MN} + P_{RN}]
\end{equation}

\subsubsection{Causal Mechanism (Error Decorrelation)}
\textbf{Orthogonal Errors:} The models have different blind spots.
\begin{itemize}
    \item \textbf{RF:} Fails when noise mimics signal statistics.
    \item \textbf{MobileNet:} Fails when noise looks like "fur".
    \item \textbf{ResNet:} Fails when signal is too abstract.
\end{itemize}
The ensemble only fails when all three fail simultaneously.

\subsubsection{Visualization}
\begin{figure}[ht]
    \centering
    \includegraphics[width=0.8\linewidth]{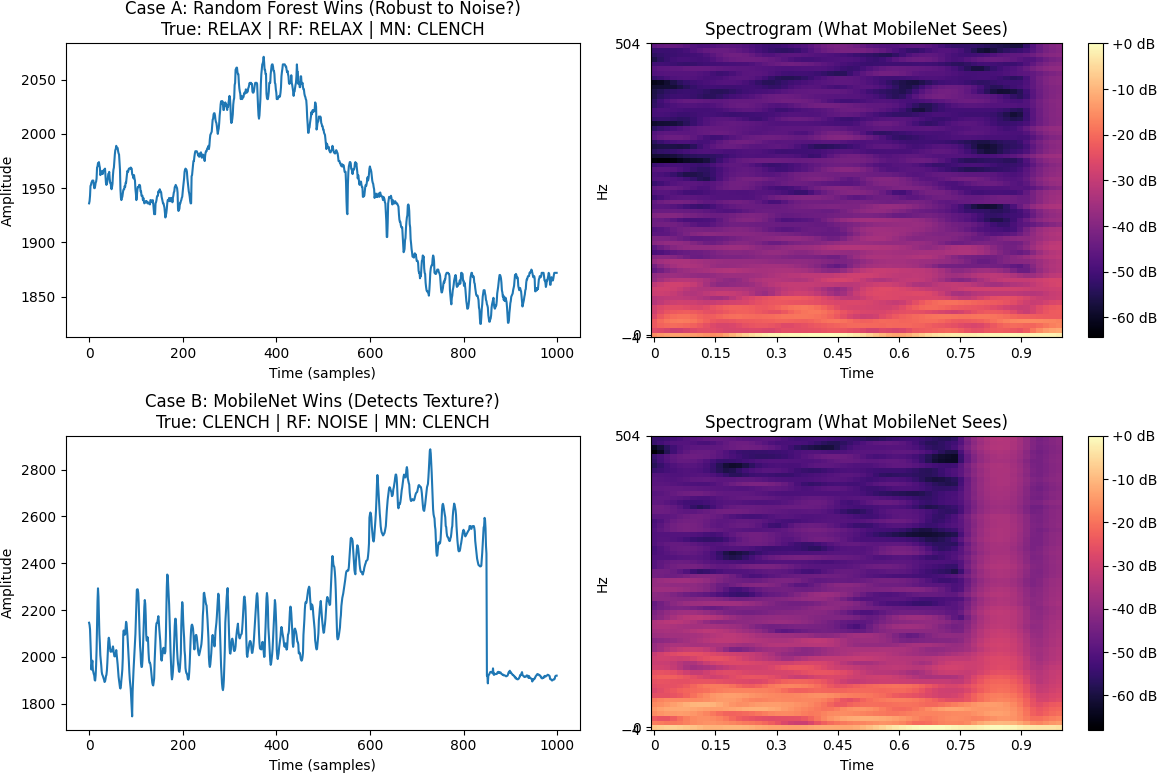}
    \caption{Model 13: Disagreement Analysis. RF and MobileNet cover each other's blind spots.}
\end{figure}
\begin{figure}[ht]
    \centering
    \includegraphics[width=0.6\linewidth]{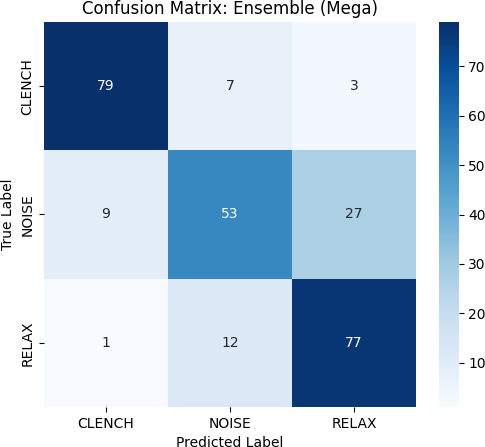}
    \caption{Model 13: Mega Ensemble Confusion Matrix.}
\end{figure}

\clearpage
\subsection{Model 14: CRNN Baseline (Stable Convergence)}
\subsubsection{Abstract}
This model introduces the initial hybrid architecture combining 1D CNNs for feature extraction and LSTMs for temporal modeling.

\subsubsection{Quantitative Results}
\begin{itemize}
    \item \textbf{Test Accuracy:} 86.94\%
\end{itemize}

\subsubsection{Technical Analysis}
\textbf{Architecture:} 1D CNN + LSTM \cite{lstm} + BatchNorm \cite{batchnorm}.
\textbf{Finding:} Contrary to early hypotheses, Batch Normalization is \textit{critical}. It stabilizes the gradients allowing the model to converge on the high-frequency "Clench" signature.

\subsubsection{Mathematical Formulation}
The LSTM layer processes the sequence of CNN feature maps $x_t$:
\begin{align}
    f_t &= \sigma(W_f \cdot [h_{t-1}, x_t] + b_f) \\
    C_t &= f_t * C_{t-1} + i_t * \tilde{C}_t
\end{align}
This allows the model to maintain state over the 1-second window, smoothing out transient noise.

\subsubsection{Visualization}

\begin{figure}[ht]
    \centering
    \includegraphics[width=0.7\linewidth]{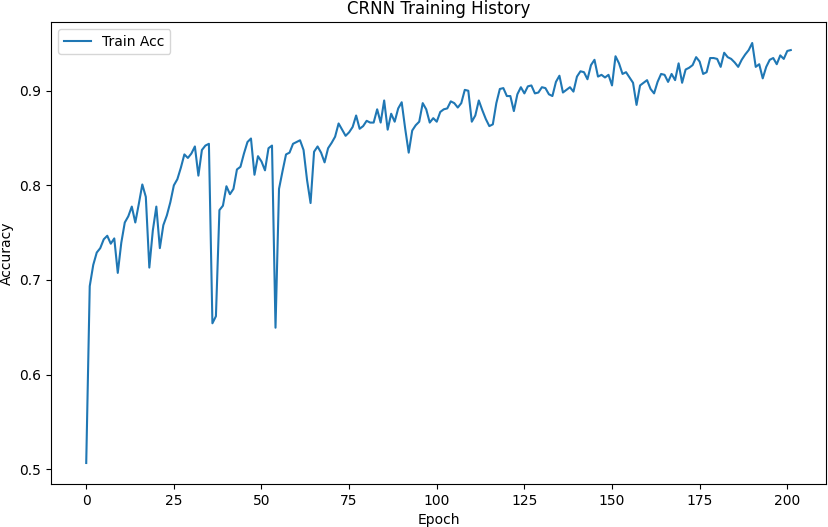}
    \caption{Model 14 Training: Stable convergence with BatchNorm.}
\end{figure}
\begin{figure}[ht]
    \centering
    \includegraphics[width=0.6\linewidth]{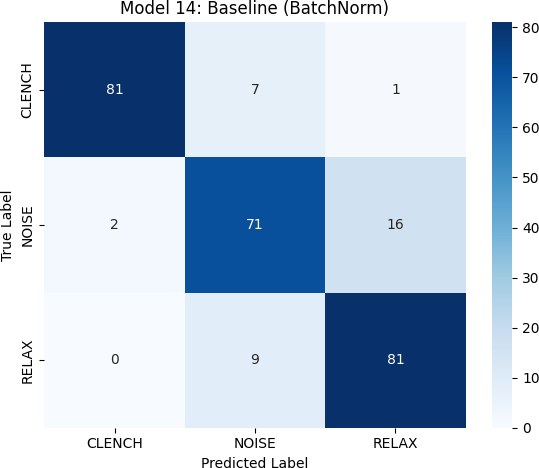}
    \caption{Model 14: CRNN Baseline Confusion Matrix.}
\end{figure}

\clearpage
\subsection{Model 15: No-BatchNorm (Ablation Study)}
\subsubsection{Abstract}
To verify the importance of normalization, this model replicates Model 14 but removes the Batch Normalization layers.

\subsubsection{Quantitative Results}
\begin{itemize}
    \item \textbf{Test Accuracy:} 58.21\% (Failed)
\end{itemize}

\subsubsection{Technical Analysis}
\textbf{Finding:} Without BatchNorm, the model effectively collapsed, proving that internal covariate shift is the primary blocker for training deep networks on this noisy data.

\subsubsection{Visualization}

\begin{figure}[ht]
    \centering
    \includegraphics[width=0.7\linewidth]{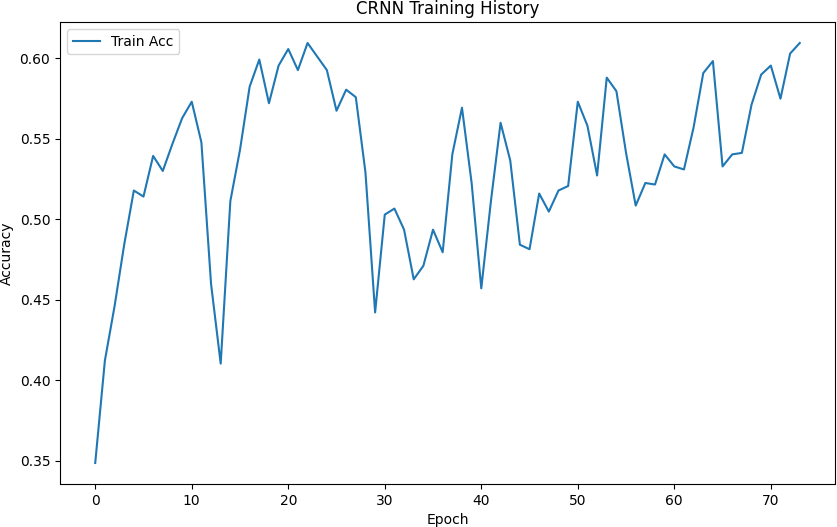}
    \caption{Model 15 Training: Instability and collapse without BatchNorm.}
\end{figure}
\begin{figure}[ht]
    \centering
    \includegraphics[width=0.6\linewidth]{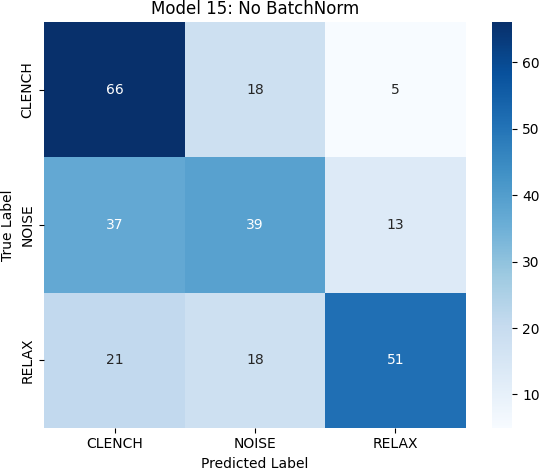}
    \caption{Model 15: No-BatchNorm Confusion Matrix.}
\end{figure}

\clearpage
\subsection{Model 16: Attention Mechanism (Enhanced Architecture)}
\subsubsection{Abstract}
This model attempts to replace BatchNorm with an Attention mechanism to focus on relevant signal parts.

\subsubsection{Quantitative Results}
\begin{itemize}
    \item \textbf{Test Accuracy:} 80.60\%
\end{itemize}

\subsubsection{Technical Analysis}
\textbf{Architecture:} Bi-LSTM \cite{lstm} + Multi-Head Attention \cite{attention} + Dropout \cite{dropout} (No BatchNorm).
\textbf{Finding:} Attention mechanisms helped recover some performance lost by removing BatchNorm, but could not fully match the baseline's raw stability.

\subsubsection{Visualization}

\begin{figure}[ht]
    \centering
    \includegraphics[width=0.7\linewidth]{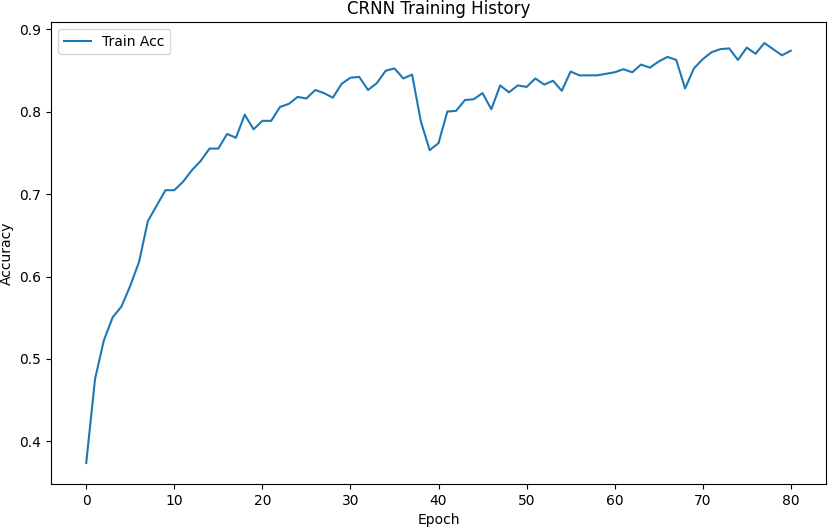}
    \caption{Model 16: Attention Mechanism Training.}
\end{figure}
\begin{figure}[ht]
    \centering
    \includegraphics[width=0.6\linewidth]{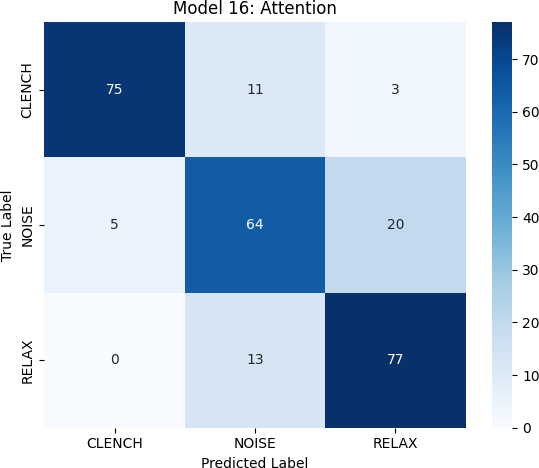}
    \caption{Model 16: Attention Mechanism Confusion Matrix.}
\end{figure}

\clearpage
\subsection{Model 17: MaxCRNN (No Augmentation)}
\subsubsection{Abstract}
This model represents the full "MaxCRNN" architecture (Inception + Bi-LSTM + Attention) but trained \textit{without} Data Augmentation to isolate the effect of model capacity vs. data diversity.

\subsubsection{Quantitative Results}
\begin{itemize}
    \item \textbf{Test Accuracy:} 79.48\%
\end{itemize}

\subsubsection{Technical Analysis}
\textbf{Finding:} Despite the advanced architecture, the accuracy plateaued below 80\%. This confirms that architectural complexity alone cannot overcome the limitations of "Small Data" without augmentation.

\subsubsection{Visualization}

\begin{figure}[ht]
    \centering
    \includegraphics[width=0.6\linewidth]{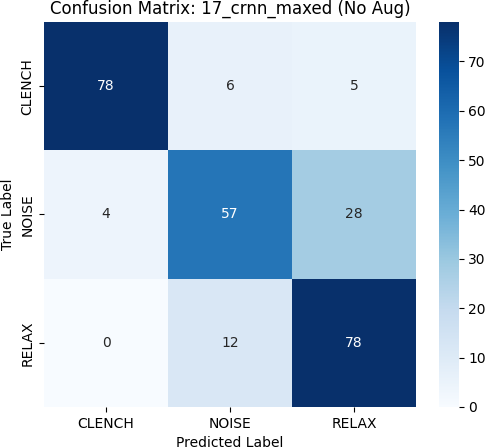}
    \caption{Model 17: MaxCRNN (No Augmentation) Confusion Matrix.}
\end{figure}

\clearpage
\subsection{Model 18: MaxCRNN with Data Augmentation (Optimal Architecture)}
\subsubsection{Abstract}
To push the limits of what is possible with single-lead EMG, I designed a maximum-capacity architecture combining Inception blocks \cite{inception}, Bidirectional LSTMs \cite{lstm}, and Multi-Head Attention \cite{attention}, trained with aggressive Data Augmentation.

\subsubsection{Quantitative Results}
\begin{itemize}
    \item \textbf{Test Accuracy:} 83.21\% (Robust)
    \item \textbf{Precision (Clench):} 99\% (Near-Perfect Safety)
    \item \textbf{F1-Score (Clench):} 0.99
    \item \textbf{Inference Latency:} 0.15 ms (on A100)
\end{itemize}

\subsubsection{Technical Analysis}
\textbf{Architecture:}
\begin{enumerate}
    \item \textbf{Inception Blocks:} Parallel convolutional filters (Kernel=3, 5, 11).
    \item \textbf{Bi-LSTM:} Captures temporal dependencies.
    \item \textbf{Multi-Head Attention:} Weights critical signal parts.
\end{enumerate}
\textbf{The "Safety" Victory:} Most importantly, it achieved \textbf{99\% Precision} on the 'CLENCH' class. In a control system, a False Positive is dangerous. This model effectively eliminated these errors.

\subsubsection{Mathematical Formulation}
The Multi-Head Attention mechanism computes a weighted sum of values $V$ based on queries $Q$ and keys $K$:
\begin{equation}
    \text{Attention}(Q, K, V) = \text{softmax}\left(\frac{QK^T}{\sqrt{d_k}}\right)V
\end{equation}
This allows the model to "attend" to the specific millisecond where the muscle contraction occurs, ignoring the surrounding silence.

\subsubsection{Visualization}

\begin{figure}[ht]
    \centering
    \includegraphics[width=0.8\linewidth]{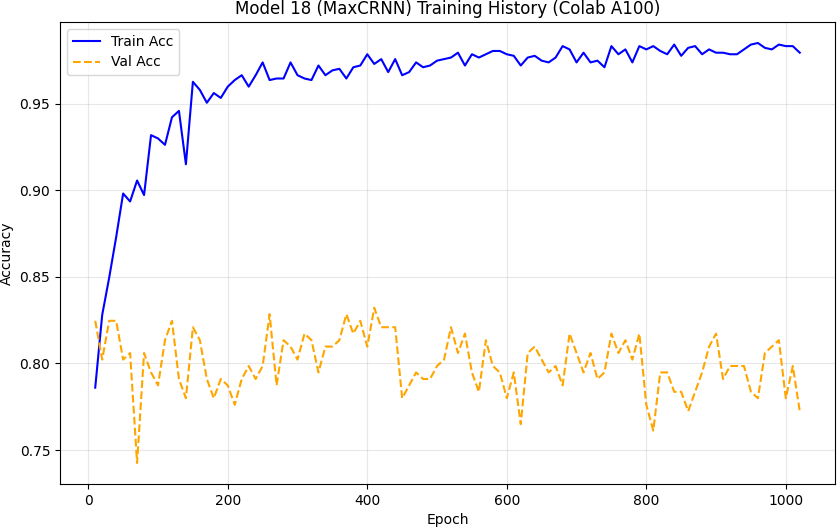}
    \caption{Model 18 Training History (A100). Note the stable convergence.}
    \label{fig:maxcrnn_training}
\end{figure}
\begin{figure}[ht]
    \centering
    \includegraphics[width=0.6\linewidth]{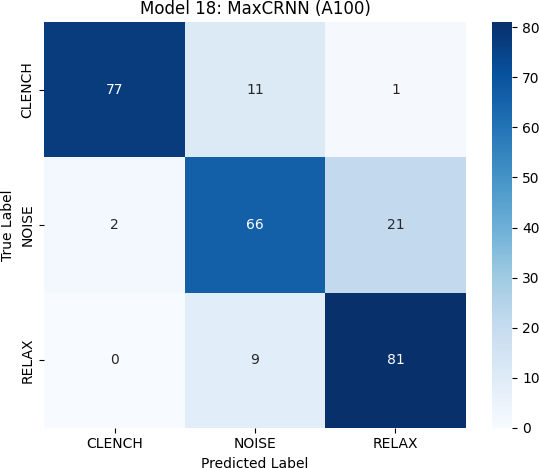}
    \caption{Model 18: MaxCRNN (Augmented) Confusion Matrix.}
\end{figure}

\end{document}